\documentclass{aa}
\usepackage{graphicx}
\usepackage{rotating}
\usepackage{txfonts}
\usepackage{natbib}
\bibpunct{(}{)}{;}{a}{}{,}
\usepackage{aalongtable}

\newcommand{\ee}{$e^\pm$}

\newcommand{\sax}{{\it BeppoSAX~}}
\newcommand{\xte}{{\it RXTE~}}
\newcommand{\wind}{{\it WIND~}}
\newcommand{\batse}{{\it BATSE~}}
\newcommand{\hete}{{\it HETE-2~}}

\newcommand{\swift}{{\it SWIFT~}}
\newcommand{\rxte}{{\it RXTE~}}
\newcommand{\xmm}{{\it XMM-Newton~}}
\newcommand{\chandra}{{\it Chandra~}}

\newcommand{\kev}{\rm keV~}
\newcommand{\flux}{\rm erg\ cm^{-2}\ s^{-1}}

\newcommand{\epc}{E_{\it peak}}
\newcommand{\eo}{E_{\it o}}
\newcommand{\xxr}{{XRRs/XRFs}}

\def\be{\begin{equation}}
\def\ee{\end{equation}}
\def\no{\noindent}
\def\ltsima{$\; \buildrel < \over \sim \;$}
\def\lsim{\lower.5ex\hbox{\ltsima}}
\def\gtsima{$\;\buildrel > \over \sim \;$}
\def\gsim{\lower.5ex\hbox{\gtsima}}
\begin{document}

\authorrunning{V. D'Alessio,L.  Piro \& E. M.  Rossi}
\titlerunning{Properties of XRRs and XRFs}

\title{Properties of X-Ray Rich Gamma Ray Bursts and X-Ray Flashes detected with BeppoSAX and Hete-2 }
\author{~V.~D'Alessio\inst{1}\thanks{E-mail:valeria.dalessio@rm.iasf.cnr.it}
\and  L.~Piro\inst{1}
\and E.~M.~Rossi\inst{2}\thanks{Chandra Fellow} }
\institute{INAF - Sezione di  Roma, Via del Fosso del Cavaliere, 100,
00113 Roma, Italy \and Max Planck Institute for Astrophysics, Garching
Karl-Schwarzschild-Str. 1
Postfach 1317
D-85741 Garching,
Germany;JILA University of Colorado 440 UCB Boulder, CO 80309-0440}
\date{Received ....; accepted ....}
\markboth{V. D'Alessio, L.Piro:Properties of XRRs and XRF }
{V. D'Alessio, L.Piro: Properties of XRRs and XRFs}
\abstract{  We study the spectrum of the prompt emission and the X-ray
and optical afterglow fluxes of 54 X-Ray Rich Gamma Ray Burst (XRRs) and
X-Ray Flashes (XRFs), observed by \sax and {\it
HETE-2}. A comparison is then performed with classical
Gamma Ray Bursts (GRBs).  The goal of this paper is to investigate the nature of XRRs/XRFs, as high redshift GRBs or off-axis GRBs,  analyzing both their prompt and  afterglow properties.
We find that the XRR/XRF spectral indexes of the Band function are similar to those of classical
GRBs, whereas the peak energy is lower by a factor of  4.  We study the optical and X-Ray afterglow properties of the
XRRs/XRFs; in particular we analyze  
the \object{XRR 011030} afterglow. We find that the X-ray and optical
flux distributions and the lightcurves of the XRRs/XRFs sample are consistent with those of classical GRBs; in
particular, they show evidence of a  break and no rising temporal slope.
We compare  these results  with the
afterglow predictions of the high redshift scenario, where XRFs are GRBs at higher
redshift  and of the off-axis scenario, where the observed differences are
due to viewing angle effects. In this last framework, we consider jets
with a homogeneous, a $-2$ power-law shaped and a Gaussian luminosity
angular distribution.
 We find that the high redshift
scenario can explain some events but not the total sample of
XRRs/XRFs. The off-axis model may be consistent with our
findings  when a homogeneous jet is considered. However,
given the uncertainties on the selection effects in our sample, a
Gaussian jet viewed at small angles from the Gaussian core and a power-law shaped cannot be
ruled out.

}

\maketitle

\section{Introduction}
  Several satellites have  observed Fast X-Ray Transients \citep{arefiev}; 
the origin of these events was attributed to a mixed contributions  from different  sources, such as flare stars and  RS CVn systems. \citet{gotthelf} first used the alternative term {\it X-Ray Flashes} for this  phenomenon. The discovery that  a large fraction of these events (in particular those with a duration  of less than 1000 $s$)   are  a class of GRBs  was made with   the Wide Field Cameras  (WFC) on \sax
\citep{heise}.

XRFs  are Gamma-Ray Bursts (GRBs)  characterized by no or
faint signal  in
the gamma ray energy range. They show an isotropic distribution on the
sky and a duration between a  few tens and $\sim 10^{3}$ seconds, like
long GRBs \citep{heise}. An intermediate class of bursts has been
observed between the XRF and the GRB classes, the X-Ray Rich Gamma Ray
Bursts (XRRs), with an X-Ray emission stronger than gamma-ray one
\citep{barr,att}. We classify  bursts  according the
definition  proposed  by \citet{lamb03}.

There are several studies of  the  spectral properties of the XRRs/XRFs;
\citet{kip03}  analyzed a sample of 9 XRFs observed by \sax  using
 untriggered
\batse data and found that
the photon indices  $\beta_1$ and $\beta_2$ of the Band function
of the XRFs are similar to those of the GRBs, instead of $\epc$, whose
value is less than 10 \kev for most XRFs. This result has been
confirmed by \citet{Sakamoto} with the analysis of 42 XRRs/XRFs
observed by \hete.

Several theories have been proposed to explain the origin of XRFs:
high redshift GRBs \citep{heise},  GRBs with a uniform
jet viewed off-axis \citep{yamaz02,yamaz03,yamazaki04}, GRBs with the Universal
Power-law-shaped jet \citep{lamb05},   a Gaussian jet
\citep{zhang03},  a ring shaped jet \citep{eichler} and a multi sub-jets \citep{toma},  a variable jet opening-angle
\citep{lamb05}, dirty  fireballs \citep{dermer},    clean
 fireballs \citep{moc}, a  photosphere dominated emission
\citep{ram} and  off- axis cannonballs \citep{dar}.

We focus here on two models: the high redshift and the off-axis
scenario.  In the former case, XRFs are high-redshift GRBs while in
the latter, they are GRBs viewed at a  large angle from the jet axis. The
main goal of this paper is to investigate the properties of XRRs/XRFs
 constrain these  theories.

The paper is organized as follows.  In \S~\ref{sec:catalogo}, we
compile a sample of 54 \sax and \hete events, catalogued as XRRs/XRFs
in the  literature, and we classify them with the same hardness spectral
ratio. This allows us to build a homogeneous sample. In
\S~\ref{sec:spectralparam}, we study and compare the distributions of
the spectral parameters $\beta_1$, $\beta_2$ and $\epc$ of the prompt
emission of XRRs/XRFs and GRBs. In \S~\ref{sec:analysis}, we describe
the observations and the data analysis of the optical and X-ray
afterglow flux; in particular we study two \chandra observations of
the afterglow of the burst \object{XRR 011030}. We present our
afterglow results in \S~\ref{sec:results} and in
\S~\ref{sec:discussion}, we discuss them in the framework of the high
redshift and off-axis scenario; in \S~\ref{sec:conclusioni} we present our   conclusions.

\section{Definition of the catalog}
\label{sec:catalogo}
We considered all the events observed until 31 December 2003 and
classified as XRRs/XRFs, available in the literature and from the web.  We
compiled a sample of 54 events, 17 observed by \sax and 37 by
\emph{HETE-2}. We have not considered the \sax bursts \object{XRF 991217}
\citep{muller} and
\object{XRF  000608} \citep{gand00}, detected only by  the WFCs,
due to the lack of spectral
information
in the literature.
 \citet{lamb03} proposed to classify bursts
according to their spectral hardness ratio $(H_h)$:  GRBs have
$ H_h=S(2,30)/S(30,400)
\le 0.32$,  XRRs  have
$ 0.32 \le H_h \le 1$ and  XRFs have
$ H_h \ge 1$  where $ S(E_1,E_2)$ is the fluence in the
energy range $ E_1-E_2$. We adopted their definition in order to build a homogeneous
sample out of a collection of events observed by different satellites.
 The \sax instruments have energy ranges compatible with the \emph{HETE-2} ones; moreover, the sensitivity of WFC,  $\sim 4\times 10^{-9}$$erg$$ cm^{-2}s^{-1}$ \citep{depasq5},  is comparable to that of WXM, equal to  $\sim 9\times 10^{-9}$$erg$$ cm^{-2}s^{-1}$ \citep{ricker02}. Thus the combined sample derived from these two satellites is homogeneous.
We calculated the hardness ratio
($H_h$) for those bursts that do not have this parameter  available, using the spectral parameters of the prompt emission. The
data are reported in Table
\ref{tab1},  \ref{tab2} and \ref{tab3}.
For completeness we also report the hardness ratio calculated in the
\sax ranges ($ H_s=S(2,10)/S(40,700)$) in Table \ref{tab3}.

All the \sax bursts include data of the WFCs (2-28 keV). At higher
energies they present data or upper limits from \sax GRBM, \batse and
\wind, except for \object{XRR 011030} and \object{XRF 020427}. The
\hete bursts include data of WXM (2-25 keV) and FREGATE (8-400 keV);
in the cases of \object{XRF 031109} and \object{XRR 031220} only the
FREGATE spectral data are available.

 Three bursts (\object{XRF 981226}, \object{XRF  990704} and
\object{XRF 020427}) have time resolved spectra: $H_h$ and $
H_s$ have been obtained as the mean weighted by the different
integration times.
For  \object{XRR 991106}
 we only have a lower limit to  the X- to $\gamma$-ray peak flux ratio
$\sim0.75$ \citep{gand99} and we  classify it as an XRR. The results are
reported in Table
\ref{tab3}.
 We
estimated typical  uncertainties of 50\% on the ratio, due to large errors on the spectral parameters.

We find that all the events, including the \sax ones, are
consistent with the XRR/XRF definition ($H_h\ge0.32$) given by
\citet{lamb03}.

The resulting total sample is of 54 bursts, 26 XRFs and 28
XRRs. We analyzed XRRs and XRFs as a unique class due to the small
number of events with X-ray and optical afterglow detections.

\section{Prompt emission: spectral  parameter distributions}
\label{sec:spectralparam}

\begin{figure}
\resizebox{\hsize}{!}{
   \includegraphics[width=5cm,angle=90]{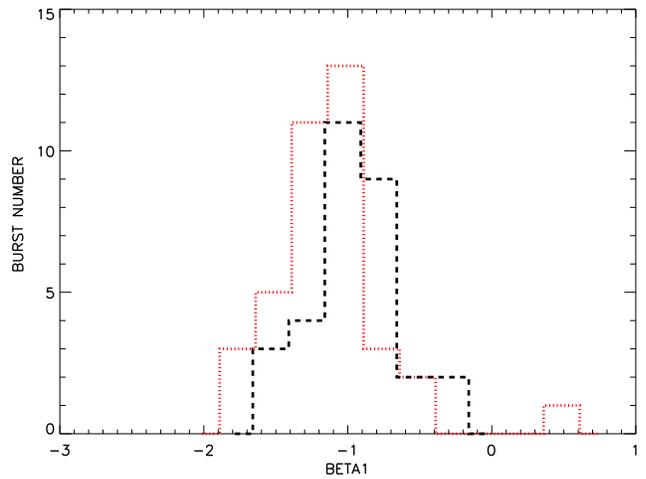}}
   \caption{Distribution of spectral slope $\beta_1$ for 38 XRRs/XRFs
   (red dotted line) and 31 GRBs (black dashed line) (see electronic
   version for color figure).}
\label{fig:alpha}
\end{figure}

\begin{figure}
\resizebox{\hsize}{!}{
   \includegraphics[width=5cm,angle=90]{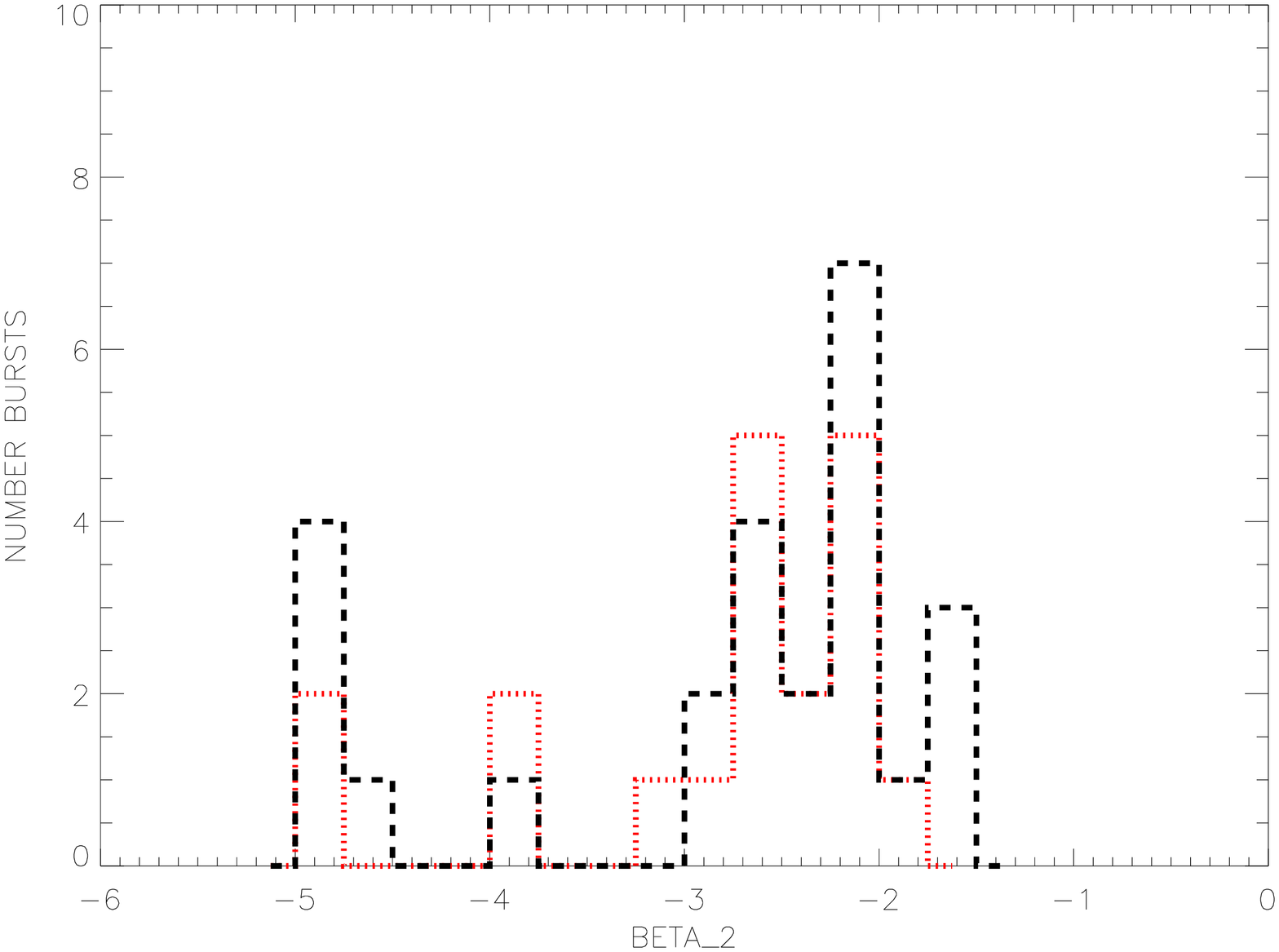}}
   \caption{Distribution of spectral slope $\beta_2$ for 19 XRRs/XRFs
   (red dotted line) and 25 GRBs (black dashed line)(see electronic
   version for color figure)}
\label{fig:beta}
\end{figure}

\begin{figure}
\resizebox{\hsize}{!}{
\includegraphics[angle=90]{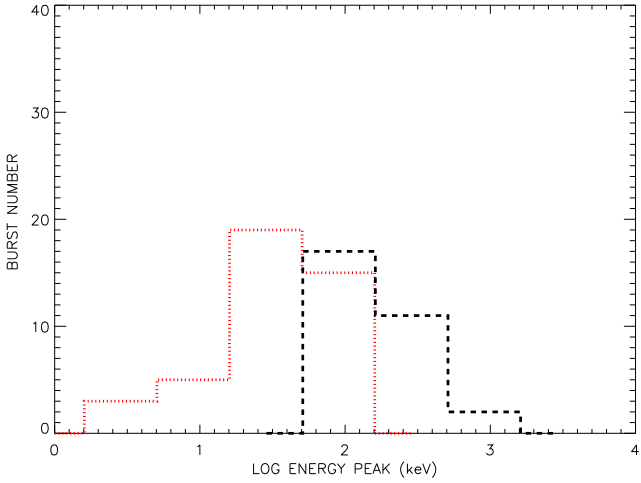}}
\caption{Distribution of the logarithm of $\epc$ for 42  XRRs/XRFs
(red dotted line) and 30 GRBs (black dashed line)(see electronic
version for color figure)}
\label{fig:ep}
\end{figure}

We studied and compared the prompt emission spectral parameters of
XRRs/XRFs and  GRBs.  We considered the events fitted by the Band
law:   $\beta_1$ ($\beta_2)$ is the  low (high) spectral index, $\epc$
is the  peak energy ($\epc =(2+\beta_1)\times \eo$) and $\eo$ is the  break
energy. In addition, we took into account
bursts described by a
power law exponential model, following the relation
N(E)=$KE^{\beta_1}\times exp(-E/\eo)$. We built up the distributions
of $\beta_1$ , $\beta_2$ and $\epc$ for  XRRs/XRFs
 (excluding parameters with no errors and upper limits)
and for 31 GRBs, 21
reported in \citet{kip03} and 10 in \citet{Sakamoto}.
 These distributions are shown in Fig.~\ref{fig:alpha},
~\ref{fig:beta}, ~\ref{fig:ep}.


The reported values for $\beta_1$, $\beta_2$ and $\epc$ are the result
of the convolution of the intrinsic distribution with the measurement
error distributions. Assuming that both are Gaussian, it is possible
to deconvolve the two distributions: we obtained the best estimate
mean value and standard deviation of the parent distribution (see
Tab.~\ref{tab10}), following the maximum likelihood method
\citep{mac98}.  We find that the mean XRR/XRF value of $\langle
\beta_1 \rangle_i \simeq -1.2$ and of $\langle \beta_2
\rangle_i \simeq -1.7$  are
 consistent with those of GRBs ($\langle \beta_1
\rangle_i \simeq -1$ and $\langle \beta_2
\rangle_i \simeq -2.3)$ within two $\sigma$.
Instead, the peak energy $\langle
E_{p,xr} \rangle_i \simeq 36$ $keV$  is significantly smaller by a factor of $\sim 4.5$ with
respect to that of GRBs ($\langle E_{p,grb} \rangle_i \simeq 162$ $keV$).

We note that 12 XRRs/XRFs have known redshift (see. Table \ref{tab7}): the mean value is $\langle
z_{xr} \rangle=1.7\pm 0.3$, with a minimum and maximum measured redshift of
$z=0.17$ and of $z=3.4$, respectively.  In order to analyse the intrinsic properties of the prompt  emission we compared the rest frame peak energy,$ \langle \tilde{E}_{p} \rangle$, of 10 XRRs/XRFs (excluding \object{XRF 020903}, having an upper limit on the peak energy and \object{XRR 030323}) and 12 GRBs. We find that $\langle \tilde{E}_{p} \rangle_i$ of XRRs/XRFs is smaller by a factor$\sim$ of 4 compared to the GRB one. In both cases,
data confirm  the \emph{soft} nature of these events.


\section{ The afterglow properties}
\label{sec:analysis}

We studied the afterglow properties of XRRs/XRFs, analyzing the
temporal profile and the distribution of the afterglow flux.  We use
the X-ray and the optical detections reported in the
\emph{GRB Coordinates Network (GCN)}
\footnote{http://gcn.gsfc.nasa.gov/} and in published papers
(see Table ~\ref{tab4} for references). In the case of \object{XRR 011030} we carried out the analysis of \chandra follow-up observations.
In Table \ref{tab4}, we list the general information for  the XRR/XRF
sample, with possible detections in the X-ray, optical and radio bands
and of the host galaxy.  In Table \ref{tab7}, we report the
redshift for 16  XRRs/XRFs.

  The events \object{XRR 971024},
\object{XRR 980128}, \object{XRR 980306} and \object{XRR 000208} have
none of this  information available in the literature and so we excluded
them from our analysis.

For 15 bursts,  X-ray afterglow observations have been
performed and all of them show an afterglow candidate. 40 bursts
present at least an optical observation. 16 of these show an optical
transient (OT) candidate, while 11  events present an  R magnitude $\le
22$ at 1 day  and are thus defined as \emph{DARK}.
10 bursts out of 20 have a radio afterglow candidate
detection. The possible host galaxy has been found for 17 bursts.

In the following section, we present the analysis of the two \chandra
observations of the \object{XRR 011030} X-ray afterglow. The study of
the lightcurves and the flux distributions of the whole sample are
 reported in \S~\ref{sec:profilo_temporale} and \S~\ref{sec:results}.

\subsection{The case of \object{XRR 011030}:   \chandra observations}
\label{sec:011030}
\begin{figure}
\resizebox{\hsize}{!}{
\includegraphics[angle=-90]{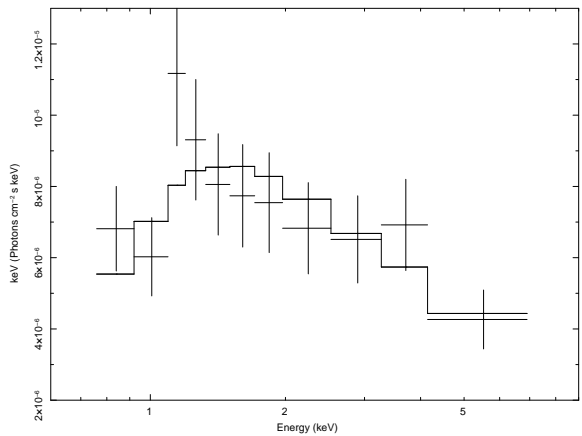}}
\caption{Spectrum of the first  observation made by ACIS-I of  \chandra
of the  afterglow  of the \object{XRR 011030} in 0.2-9 \kev}
\label{011030}
\end{figure}

\object{XRR 011030} was  detected by WFC1 of \sax on October 2001
\citep{gand}. The duration of this burst is 1200 $s$ in 2-28 \kev
with a total fluence of $S_x$=1.2$\times10^{-6}$ erg cm$^{-2}$ \citep{galli}. The
 spectrum is well fitted by a power-law with a photon index $
\Gamma=-(1.8\pm0.2)$ \citep{galli}. Two X-ray observations have
been performed for this event, after 11 and 31.2 days, that lasted 46.61 and 20.12 $ks$ respectively; they  were
 made by ACIS-I on \emph{CHANDRA}.  A new
X-ray  transient was  discovered  at R.A. 20:43:32.5 and DEC +77:17:17.4,
 at 1.2 arcsec from the radio transient position \citep{harr}. We
processed  the   \chandra data  available in the archives of \chandra
observations \footnote{http://cda.harvard.edu/chaser/mainEntry.do} with CIAO
version 3.1, using the task \emph{acis\_process\_events}. The spectra have
been extracted selecting  a circular  area  around the  source, which is
in the center,   to optimize the signal to noise ratio;  the spectra
of the background have been extracted  from a larger circular region without
sources. We performed  the spectral analysis  with XSPEC version 11.3.1.
The spectrum of the first X-ray observation  is well described  by an
absorbed power-law  with $N_H$=2.96$^{+0.60}_{-0.65}\times10^{21}$
cm$^{-2}$,  photon index $\gamma=1.72^{+0.19}_{-0.20}$ and
$\chi_{\nu}^{2}=0.76$
with 9 d. o. f.. The flux in 2-10 \kev is $F=(5.75\pm0.86)\times10^{-14}$ $\flux$.
The spectrum of the first X-ray observation   in the energy range between
0.2-9 \kev is
shown in Fig  \ref{011030}.

In the second observation there are too few counts and we fitted its
spectrum with an absorbed power-law with $N_H$ and $\gamma$ kept  at
the value previously found.  Thus the flux at  2-10 \kev is
$F=(5.58\pm1.45)\times10^{-15}\flux$.  We obtained a temporal decay
index of $\delta_x=-(2.25\pm0.60)$ between the two \chandra
observations.  We note that our results are in agreement with those
calculated by \citet{sako05} and \citet{gen05}.

\subsection{ The lightcurves}
\label{sec:profilo_temporale}

\begin{figure}
\resizebox{\hsize}{!}{
\includegraphics[angle=90]{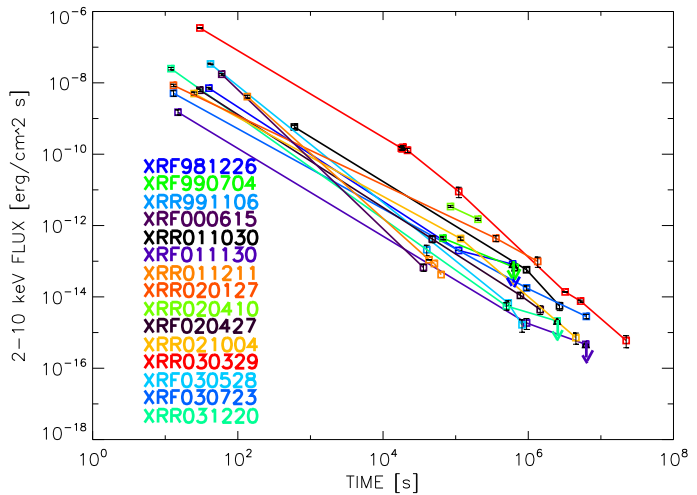}}
\caption{X-ray  light curves  of 15  XRRs/XRFs with   afterglow
observations in the
range 2-10 \kev (the arrows  indicate  the upper limits); the points are connected with straight lines.
For event \object{XRR 991106} we used the afterglow flux  by \citet{depasq}. For completeness we report also the prompt data (see electronic version for color figure).}
\label{light_curve}
\end{figure}

The temporal profile of the X-ray afterglows are studied
collecting all the available ToO observations made by
\emph{BeppoSAX},
\chandra and \xmm. The detection epochs range from a 
 few hours to more than a month.  We obtained
 a sample of 15 bursts: 9 events have observations within 1 day and
 the other 6 events have later observations.  We report in
 Fig. \ref{light_curve} the prompt and the afterglow X-ray data for 
 the XRR/XRF sample.

To detect  possible temporal changes in the X-ray lightcurve,
we separately considered the decay indexes of the two previously
defined sub-samples of events.
For the 9 events with early-time observations,
we found a weighted mean value of $\langle\delta_{x}\rangle$=$-(1.0\pm0.1)$. For the other sample (excluding two events
with upper limits and including the late \xmm observation of
\object{XRR 030329})  we found instead
 $\langle\delta_{x}\rangle$=$-(1.5\pm0.1)$.  The early and late time
 decay indexes are not compatible at a  5 $\sigma$ level. This suggests the
 presence of breaks in the afterglow evolution as commonly
 observed for GRBs.
 

In particular, \object{XRR 030329} and \object{XRR
 021004} show a temporal decay index  of
 $\delta_{x}$=$-(1.9\pm0.2)$ between $ 111-3222$ $ks$ and
 $\delta_{x}$=$-(1.8\pm0.5)$ between $ 12-4500$ $ks$.  These values are
 consistent with the decay index expected after the jet break, where 
 jet matter expands laterally at a velocity close to the speed of
 light \citep{rhoads99}.
 The presence of an achromatic break for \object{XRR 030329} is also
supported by optical data \citep{tiego}. A  possible
break in the optical lightcurve was observed in 4  other  cases:
\object{XRR 010921}
 \citep{pricec}, \object{XRR 011211} \citep{holland01}, \object{XRR
 020124} \citep{bloomd} and \object{XRR 030725}
 \citep{pugliese}. Thus, the presence of a break in the afterglow
 lightcurve of XRRs/XRFs seems to be, as  in GRBs,  a common feature.

We now focus on the early-time ($<1$ day) lightcurve.  The X-ray
afterglow data always show a decreasing temporal profile.  In two cases,
(\object{XRR 030329} and \object{XRR 011211}), we have data as early as a few hours
(5 and 10 respectively) after the trigger.
 Likewise, the optical lightcurves always show a fading behavior, sometimes
together with a plateau transition
(\object{XRR 021004},  \citet{mirabal} and \object{XRR 030723},
\citet{huang04}). The earliest data point is at $\sim 0.3$ $hr$ for
\object{XRR 021004}.   The XRR/XRF X-ray and optical lightcurves never show evidence of a rising
temporal behavior. If present, this should occur early as
 0.3  $hr$ after the explosion.

\subsection {The X-ray and optical flux distributions}
\label{sec:results}

We analysed the X-ray and optical flux distributions of XRRs/XRFs,
and we compared them to the GRB ones.  Ideally, we would like to compute and
compare the luminosity distributions of the two classes. This is
hampered  by the paucity of bursts with known redshift in our
sample. However, if the two populations have the same redshift
distribution (as in the off-axis model) or the distributions do not
overlap appreciably (as in the high-redshift scenario), the flux
comparison is meaningful.  We also used the available redshift
measurements to further test our results.

In order to test the off-axis model, we studied the afterglow
flux  in 1.6-10 \kev band
at $40$ $ks$ ($F_{x}$).  At early times 
the three jet model  lightcurves are strongly dependent on the viewing angle
 and they  bear distinctive characteristics for each model.

For the analysis, we only used the subsample of 9 early afterglow
events.  This reduced the contribution from the presence of different
decay slopes to the dispersion in the flux distribution. Moreover, for
this sample the extrapolated flux $F_{x}$ is more robustly
constrained.

\begin{figure}
\resizebox{\hsize}{!}{
\includegraphics[angle=90]{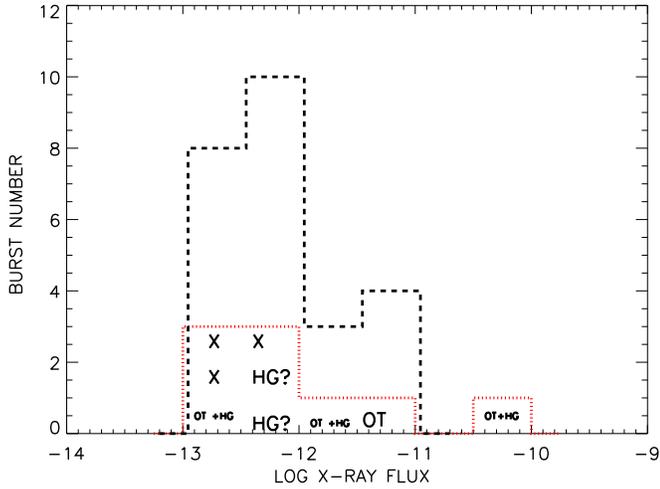}}
\caption{Distribution of the logarithm of X-ray  flux at 40 $ks$ in units of  $\flux$  for 9  XRRs/XRFs (red dotted line) and 25  GRBs (black
dashed line). OT=XRRs/XRFs with
optical transient, HG=XRRs/XRFs with host galaxy, X=XRRs/XRFs without HG and OT (see
electronic version for color figure)}
\label{flussox}
\end{figure}

We calculated the mean value and the standard deviation of the total
and parent distributions of $F_{x}$, according to the likelihood method.
We compared these results with those found with the same analysis for
a sample of 25 GRBs by \citet{depasq}. Our results are reported in
Table \ref{tab12}.
The resulting distributions are shown in Fig.\ref{flussox}. The mean
ratio between the GRB and the XRR/XRF flux is $\langle
HR_{x}\rangle_i=0.9\pm1.2$.

\begin{figure}
\resizebox{\hsize}{!}{
\includegraphics[angle=90]{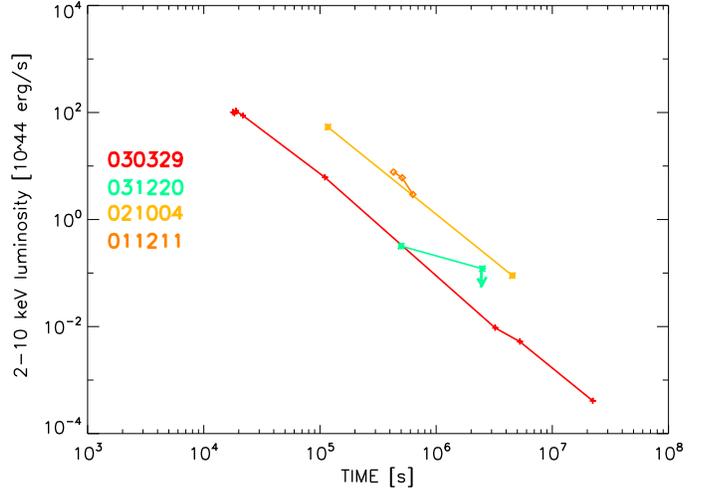}}
\caption{X-ray light curves of XRRs/XRFs with known redshift  in the rest
frame of the sources in units of $10^{44}erg$, compiled using  $\chandra, \sax,
\xte, \xmm$ observations of the afterglow. The arrow  indicates   upper limit (see electronic version for color figure).}
\label{curve_l_x}
\end{figure}

Then, we calculated the X-ray Luminosity at $40$ $ks$ for the
XRRs/XRFs with known redshift,

\begin{equation}
L(\nu,t)=F_{\nu}(\nu,t) 4\pi D^2(z)(1+z)^{1-\alpha+\delta_x},
\label{eq:flux}
\end{equation}
\citep{lamb00}, where $\alpha$ is the spectral energy index, $\delta_x$ is the
temporal decay index, $ L_{\nu}(\nu,t)$ is the luminosity at the time
$t$ and at the frequency $\nu$ and $D(z)$ is the comoving distance. We
found 4 events with known redshift, 3 of them with an early X-ray
observation.  The light curves are plotted in Fig.\ref{curve_l_x}.
We adopt a cosmology $\Omega_{\Lambda}=0.7$, $\Omega_{M}=0.3$, $H_0=70$ $km$$ s^{-1}Mpc^{-1}$, $q_0=-0.55$, $k=0$.
The mean luminosity (units of $10^{44}erg s^{-1}$) is $\langle
logL_{x}\rangle_i$=0.6$\pm0.1$, which favorably compares with that of
the \citet{depasq} sample of 12 GRBs: $\langle
logL_{x}\rangle_i$=0.5$\pm0.2$.

For the optical flux, we used OT observations at different times.
 We corrected
 the $R$ magnitude, $m_R$,  for galactic extinction, following the calculations presented by
\citet{depasq}.
 We extrapolated the flux at $40 ks$ ($F_{o}$)
 using the temporal decay index $\delta_o$ of the OT, when 
 available, and a value of $\delta_o=1.15$ otherwise
 \citep{depasq}.
Our results are
 reported in Table \ref{tab5}.  For \object{XRF 020903} and
 \object{XRR 011091} the flux extrapolated at $40$ $ks$ from the OT
 observations is overestimated  compared to the observed upper limits.
 It may be due to a  possible break in the light curve: in these cases we
 calculated $F_o$ using the upper limits, measured respectively at
 86.4 $ks$ and 79.5 $ks$.

\begin{figure}
\resizebox{\hsize}{!}{
\includegraphics[angle=90]{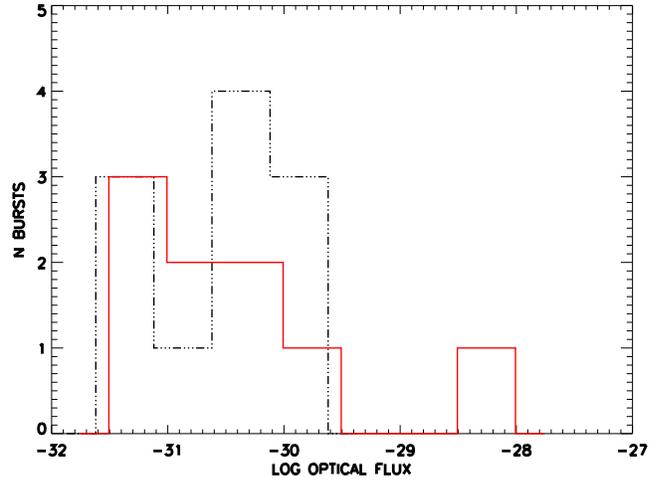}}
\caption{Distribution of the logarithm of optical  flux at 40 $ks$ in units of $W$$m^{-2}$$Hz^{-1}$  for 9  XRRs/XRFs (red dotted line) and 11  GRBs (black
dashed line). See
electronic version for color figure.}
\label{flusso_ottico}
\end{figure}

We found 9 XRRs/XRFs with an optical
detection within 1 day. We compared them with
11 OTGRBs (GRB with OT).   The ratio between the OTGRBs and the
XRRs/XRFs is $\langle HR_{o} \rangle=0.9\pm1.2$.
Both the optical and X-ray flux
distributions,  shown in Fig. \ref{flusso_ottico},  are similar between GRBs  and XRRs/XRFs.

We also analyzed the optical to X-ray flux correlation for both OT and
$DARK$ XRRs/XRFs. We find
$\langle log(f_o/f_x) \rangle$=$0.3\pm0.3$ and $\sigma=0.9$ for XRRs/XRFs and $\langle log(f_o/f_x) \rangle$=$0.4\pm0.1$ and $\sigma=0.6$ for GRBs.The ratio  is compatible within 1 $\sigma$ between the
two  classes. Instead, the standard
deviation of the XRRs/XRFs distribution is greater than a factor
$\sim$ 3 compared to the GRBs , even if we remove the three bursts
\object{XRF 981226},
\object{XRF 990704} and
\object{XRR 020410},  which have a
value of the optical to X-ray  flux ratio smaller than the minimum value
found for the  GRB sample.

We calculated the ratio between the afterglow X-ray flux (at
40 $ks$) and the prompt gamma-ray flux (40-700
$keV$). We considered 8 XRRs/XRFs (with early observations) and 9 GRBs
\citep{depasq5}. We found for the XRR/XRF and GRB parent distributions
 $\langle log(F_{x}/F_{\gamma})\rangle_i$=1.35$\pm0.33$,
$\sigma=0.61^{+0.28}_{-0.15}$ and $\langle
log(F_{x}/F_{\gamma})\rangle_i$= 0.39$^{+0.18}_{-0.24}$,
$\sigma=0.56^{+0.26}_{-0.15}$, respectively.
  These value are not consistent at the 3 $\sigma$ level.


\section{Discussion}
\label{sec:discussion}

\subsection{The  high redshift scenario }

In this section, we test the possibility that the observed global
properties of the XRFs are {\it only} due to a distance effect. The
population of XRRs/XRFs is therefore assumed to have the same {\it
intrinsic} properties of the GRBs but a higher average  redshift.
This scenario would naturally explain the spectral parameter
distributions of the prompt emission of XRRs/XRFs vs GRBs: since
 XRRs/XRFs are on average more distant, the
observed average spectrum is rigidly red-shifted.
From the ratio of the observed peak energies, assuming
$\langle z_{grb} \rangle= 1$ for GRBs, we
estimate $\langle z_{xr} \rangle \approx 8$.

Under this assumption, the
X-ray afterglow at a given time  would appear dimmer. Assuming
the same spectral and temporal slopes ($\alpha=-1.1$
and $\delta=-1.3$ ) \citep{depasq} and
$\langle z_{xr} \rangle=8$, Eq.~\ref{eq:flux} yields an  X-ray flux ratio
of $\sim 12$.  From our analysis we obtained $\langle HR_{x}
\rangle=0.9\pm1.1$.
 
\begin{figure}
\resizebox{\hsize}{!}{
\includegraphics[angle=90]{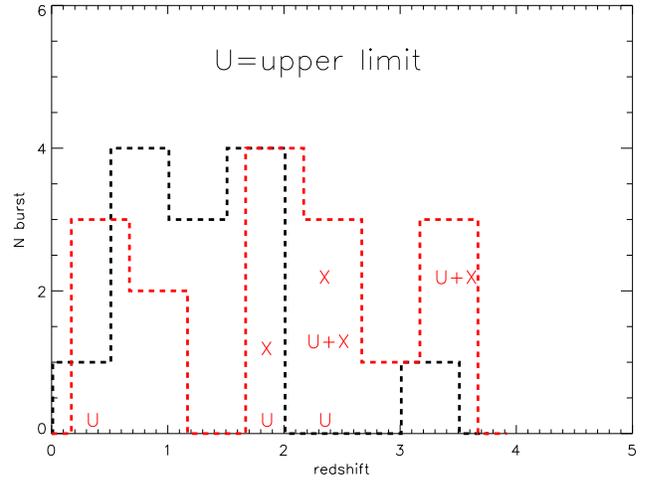}}
\caption{Distribution of the redshift   for 16  XRRs/XRFs (red dashed line) and 13  GRBs (black
dashed line). U= XRR/XRF with upper limits, T=XRR/XRF with early  X-ray afterglow detection (see
electronic version for color figure)}
\label{redshift}
\end{figure}

 In order to estimate bias effects  due to  distance, we calculate the prompt X-Ray flux for a typical GRB with $E_{iso}=10^{53}erg$,
 spectral parameters $\beta_1=-1$, $\beta_2=-2.3$ and $\epc=300$ $keV$ (value obtained using the Amati relation), at $z=8$. We  obtain an observed flux in 2-10 $keV$ (assuming a burst duration of 10 $s$) of $F_l=1.2$ $10^{-8}ergcm^{-2}s^{-1}$, $\sim$ 3  and 1.3 times greater then the sensitivity of the WFC and WXM respectively. Thus we expect  no relevant selection effects. They  appear only for GRBs with $ E_{iso} \le 10^{53}erg$ and  $z\ge8$.
 In order to test the high redshift  model, we compare the distribution of the redshift for 13  GRBs and 12 XRR/XRF of our sample, Fig. \ref{redshift}; the two distribution are similar with a mean value of $\langle z_{xr}
\rangle=1.7\pm0.3$ and $\langle z_{grb}
\rangle=1.3\pm0.2$, compatible within 1 $\sigma$. Moreover, the probability that they belong to the same parent population is $p=0.15$. Even if we consider the subsample of events  analysed for the X-ray afterglow,we find compatible  values: $\langle z_{xr}
\rangle=1.6\pm0.7$ and $\langle z_{grb}
\rangle=1.5\pm0.4$ Thus this result  suggests  that the XRRs/XRFs  and GRBs  have  similar redshift. 

We also compare  the rest frame  energy peak for XRRs/XRFs and GRBs;  we find that there are several  event XRR/XRF with an intrinsic Peak energy significantly lower then  to the GRB one. However there are also some events, like \object{XRR 020124} and  \object{XRR 021004}, with an $E_p$ in the rest frame consistent with that of  GRBs.

For high redshift objects ($z \geq 5$) no optical afterglow is
expected: the Lyman-$\alpha$-forest completely  absorbs  the emission
in the optical range \citep{fruch}.  We find that 7  XRR/XRF out of
40 are \emph{DARK} and show neither a  candidate  host galaxy nor redshift estimation. Thus some of them  could  be at high redshift. We calculated the pseudo-redshift for these bursts (see Table ~\ref{tab7}). Excluding \object{XRR 021112} with a pseudo-redshift  $pz=4.62\pm4.33$, all the other bursts  have $pz$  less than 1.2. If that estimation holds true, they are probably \emph{DARK} because of absorption \citep{depasq, jakobsson}.
In conclusion, the total sample of the XRRs/XRFs seems to be not
compatible with the high redshift model, even though we cannot
exclude that some  of them could be GRBs at $z \geq 5$. This fraction
can be under represented in our sample, if is biased toward
brighter (i.e. closer) objects.

 \subsection{The off-axis scenario}
 
 In this section, we discuss the viability of the ``off-axis
 scenario'' to explain GRB, XRF and XRR dissimilarities as due to
 differences in the observer line of sight. We specifically consider
 three models. First, the Universal Power-law-shaped (UP) jet \citep{lamb05}. Then, the Quasi-Universal Gaussian (QUG)
 jet \citep{zhang03} and  the off-axis
 Homogeneous (OH) jet model \citep{yamazaki04}.

The three jet structures differ for the distribution of kinetic energy
per unit solid angle $dE/d\Omega$ across the jet surface. In the UP
jet model, the jet is boundless ($\theta_{j}=90^{\circ})$ and the
energy distribution, outside a core angle $\theta_{c}$, follows a
power-law with index $ -2$ \citep{rossi02,zan}. In the QUG jet,
$dE(\theta)/d\Omega=(dE/d\Omega)_{0}\times e^{-\theta^{2}/(2\theta_{k}
^{2})}$, shows a nearly constant energy core ($0<\theta \le\theta_k$)
and an exponential decrease for $\theta >
\theta_k$. In the uniform jet model, $dE/d
\Omega$ is constant within the aperture ($\theta_{j}$) of the
jet and it drops sharply to zero outside.
 All models
 assume the Amati relation \citep{amati02} extended to XRF peak
 energies \citep{lamb03}; it allows us  to relate the isotropic
 equivalent energy $E_{iso}$ to the rest frame peak $\tilde{E}_{peak}$
 in the $\nu F_{\nu}$ spectrum of the prompt emission,

\be
\tilde{E}_{peak}\sim C \,\,{\rm keV} \left(\frac{E_{iso}} {10^{52} {\rm ergs}} \right)^{0.5},
\label{eq:amati}
\ee

\no
where C follows a log-normal distribution with best fit parameters
$\langle C \rangle=90.4$ and $\sigma_C=0.7$ \citep{lamb05}. In the
UP and QUG jet models there are not specific assumptions or
predictions for the shape of the spectrum as a function of the viewing
angle; instead, the spectral slopes are expected to be angle
independent in the OH jet model. This is consistent with our observed spectral
parameter distributions (Figs.~1 and 2).

The afterglow predictions and data are compared in the following
discussion. Here we only consider the sub-sample of events used for
the X-ray flux analysis in \S4.3 and their $\langle E_{p,xr}\rangle$
reported in Table \ref{tab10}.   We proceed as follows.  We define the average
parameters of the jet qualifying a XRR/XRF vs GRB  for each jet structure. XRR is
defined by an observed hardness ratio of $H>0.32$.  It corresponds to
a peak energy $E_{peak} \lsim 80$ keV, if we fix the spectral slopes
to $\beta_1=-1$ and $\beta_2=-2$ in the Band spectrum. Thus, in the
QUG and UP jet model, we can evaluate the viewing angle $\theta_o$ at
which the transition between GRBs and XRRs occurs. In the OH jet model
is $\theta_o=\theta_j$.  Then, we deduce the average viewing angle for
GRBs and \xxr~ in our sample from the observed average peak energies:
$\langle \tilde{E}_{p,grb}\rangle\simeq 210\,(1+z)=410$ keV and
$\langle \tilde{E}_{p,xr}\rangle\simeq 68\,(1+z)=136$ keV (assuming
$\langle z \rangle=1$). Simulations of X-ray afterglow light curves 
allow us an estimate of the expected GRBs over \xxr~ flux ratio at
$40/(1+\langle z\rangle)=20$ $ks$  for our sample. We note here that the average
$E_{peak}$s used in this discussion are obtained from a subsample of 7
out of 9 XRRs/XRFs and of 14 out of 25 GRBs; the associated
uncertainties are $ 10\%$ (see Table 4).  The uncertainties of the
expected flux ratios are consequently at least of $10\%$.  We
comment on how the model predictions compare with our result of a flux
ratio of the order of unity (\S4.3).

The parameters adopted by \citet{lamb05} for the UP jet are:
$\theta_c=0.26^{\circ}$, $\theta_j=90^{\circ}$ and $E_{iso}(0)=2\times
10^{54}$ ergs and $E_{iso}\simeq\frac{E_{\gamma}} {29} \theta^{-2}$,
where $E_{\gamma}\simeq 1.2\times 10^{51}$ erg is the total standard
energy. The off-axis angle is related to the peak energy by
$$\frac{\theta}{\theta_c}=\left(\frac{\tilde{E}_{peak}(0)}{\tilde{E}_{peak}}\right).$$
This implies that for $\frac{\theta_o}{\theta_c}>8 \simeq 2^{\circ}$
the observer detects \xxr. The mean viewing angles for GRBs and
XRRs/XRFs in our sample are $\frac{\theta_{grb}}{\theta_c}\simeq 3.1$
and $\frac{\theta_{xr}}{\theta_c}\simeq 9.4$ respectively.  The
expected X-ray flux ratio is {\bf $\simeq 0.5$}.  However, we
note that using the parameters adopted by \citet{lamb05}, it may be
possible to explain the large range of $E_{iso}$ needed to account for GRBs and XRFs
but not the time of the break. Due to the small core angle, we would expect the afterglow break
to occur on average before $40$ $ks$. This is not, however, what we observe
(see \S4.2 and Fig.5). If, instead, we use the  standard geometrical relation between $E_{iso}$ and $E_{\gamma}$,
we get $\theta_c=\left(\frac{2 E_{\gamma}}{E_{iso}}\right)^{0.5}\simeq 2$ degree.
The break in the lightcurve is thus expected around $\sim$ few days (Fig. \ref{jetinomogeneo}), 
 in better agreement with our results in \S4.2. In this case, the expected X-ray flux
ratio is $\simeq 20$.

\citet{zhang03} constrain with data the average parameters of the QUG jet:
 $\theta_k\simeq 5.7$ degree and a standard total energy of $E_{\gamma}\simeq 1.3 \times 10^{51}$
 erg. This implies an average $E_{iso}(0)\simeq 2.6 \times 10^{53}$
 erg. Since
$$\frac{\theta}{\theta_{k}}=\sqrt{4\,\log{\frac{\tilde{E}_{peak}(0)}
{\tilde{E}_{peak}}}},$$ XRRs/XRFs are detected for $\theta_o \gsim 2.1
\theta_k$.  In our sample, we detect GRBs observing at an average
angle of $\frac{\theta_o}{\theta_k}=0.7$ and XRRs/XRFs at
$\frac{\theta_o}{\theta_k}=2.2$.  Thus, the expected X-ray flux ratio is 
$\sim 11$ (Fig. \ref{jetgaussiano}).

In the off-axis homogeneous jet model, a GRB is detected for
$\theta_o \le\theta_j$. \citet{yamazaki04} assume a power-low
distribution of opening angles and a log-normal distribution for
$E_{\gamma}$ with $\langle E_{\gamma}\rangle \simeq 1.2 \times 10^{51}$ erg.
The parameters of the ``average'' homogeneous jet
corresponding to the observed events in Fig.~\ref{flussox} are: $
E_{iso}=2.1 \times 10^{53}$ erg and $\theta_j \simeq 6.1$ degree, where
we have used Eq.~\ref{eq:amati} and
$\frac{1}{2}\theta_j^{2}E_{iso}=E_{\gamma}$.   In this model,

$$\frac{\theta}{\theta_j}=1+\frac{1}{\theta_j\Gamma}\sqrt{\left(\frac{\tilde{E}_{p}(0)}
{\tilde{E}_{peak}}-1 \right),}$$ 

where $\Gamma$ is the Lorentz
factor.  From the
observed energy peak ratio, we derive
$\frac{\theta_{xr}}{\theta_{grb}}\simeq 1.03$ (with
$\Gamma=500$).
Correspondingly, we expect a X-ray flux ratio  of $\sim 1.1$  (Fig. \ref{jetomogeneo}).


 In summary, the OH predictions are in best agreement with data.
The conclusions for the UP jet depend instead on the assumed core size, 
which is still a poorly constrained parameter of the model. If we chose 
the size core to match the large spread of $E_{iso}$,the UP also   
favourably compares with the data.

\begin{figure}
\resizebox{\hsize}{!}{
\includegraphics[angle=0]{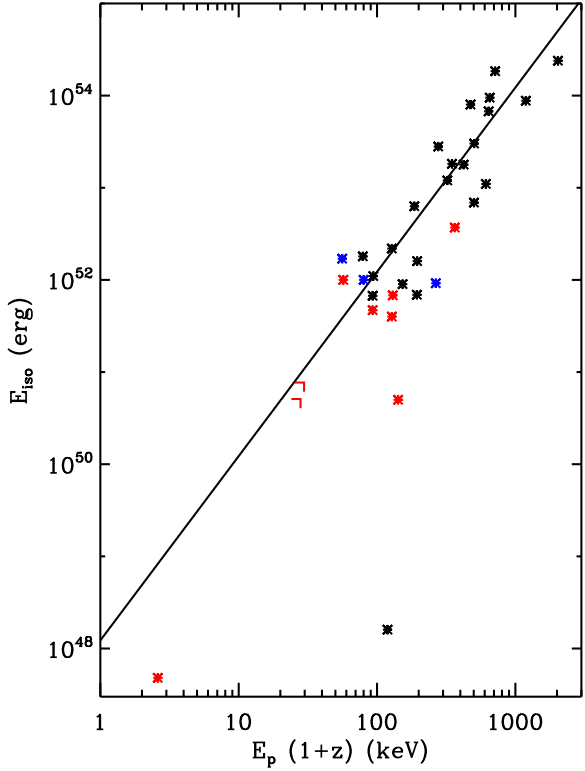}}
\caption{Amati relation   for 12  XRRs/XRFs (red and blue point) and 22  GRBs (black
point) from \citet{ghir}. The blue point are XRRs/XRFs with early  afterglow observation and known redshift. The right angles  are the upper limits. (see
electronic version for color figure).}
\label{amati}
\end{figure}

\begin{figure}
\resizebox{\hsize}{!}{
\includegraphics[height=10cm,width=7cm,angle=90]{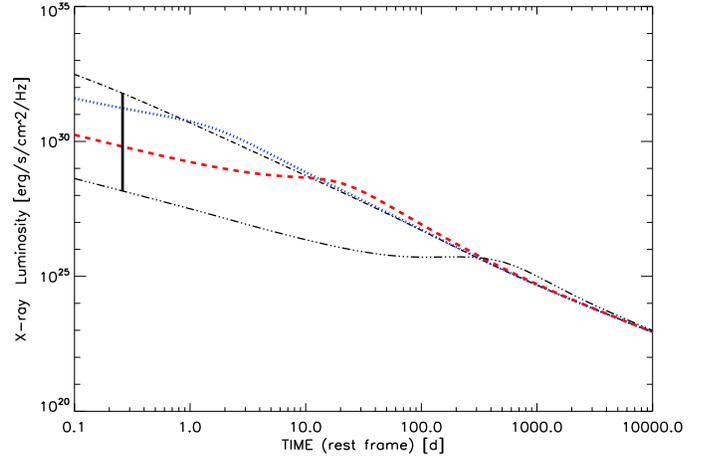}}
\caption{Light curves for  a power-law shaped   jet (UP) observed from different viewing angles.
From top to bottom: $\theta_o$=0, 3.1, 9.4, 36 $\theta_c$, where $\theta_{c}=2^{\circ}$. The estimated  viewing angle for a GRB is
$\theta_{o,grb}$=$3.1\theta_{c}$ (blue dotted line) and for an XRR/XRF is  $\theta_{o,xr}$=$9.4\theta_{c}$ (red dashed line) (see electronic version for color figure). The other parameters are:
  $\theta_{j}=90^{\circ}$ ,
$\Gamma_{0}=500$, rest frame  frequency $=10\,keV$, $\epsilon_{b}=0.01$,
 $\epsilon_{e}=0.1$, z=1, p=2.5, n=10 $ cm^{-1}$.  The straight line marks the time ($20\,ks$)
 when  we extrapolated the observed
fluxes.}
\label{jetinomogeneo}
\end{figure}

\begin{figure}
\resizebox{\hsize}{!}{
\includegraphics[height=10cm,width=7cm,angle=90]{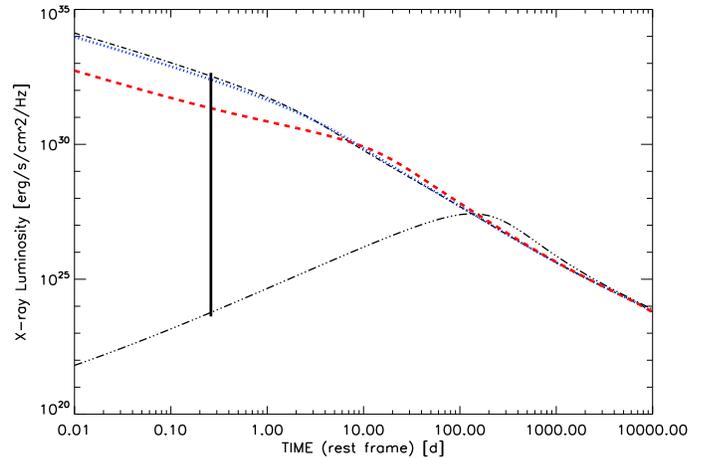}}
\caption{Light curves of a Gaussian  jet (QUG)  observed from different viewing angles  using the model described
by \citet{zhang03}. From top to bottom: $\theta_o$=0, 0.7, 2.2, 8.0 $\theta_k$, where $\theta_{k}=(5.7^{+3.4}_{-2.1})^{\circ}$. The estimated viewing angle for a GRB is
$\theta_{o,grb}$=$0.7 \theta_{k}$ (blue dotted line) and for an XRR/XRF is  $\theta_{o,xr}$=$2.2\theta_{k}$ (red dashed line) (see electronic version for color figure).  The other parameters are: $\theta_{j}=90^{\circ}$,   $\Gamma_{0}=500$,
rest frame frequency $=10\,keV$, $\epsilon_{b}=0.01$, $\epsilon_{e}=0.1$,
z=1, p=2.5,  n=10 $ cm^{-1}$.
 The straight
line marks  the time ($20\,ks$) when  we extrapolated the observed
fluxes.}
\label{jetgaussiano}
\end{figure}

\begin{figure}
\resizebox{\hsize}{!}{
\includegraphics[angle=90]{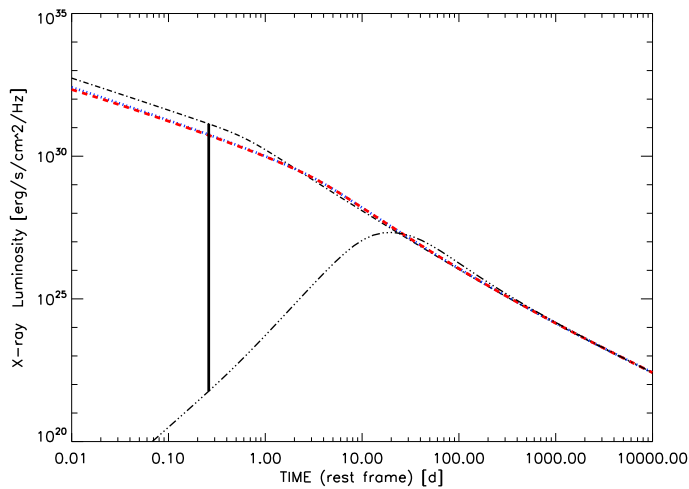}}
\caption{Light curves  of an homogeneous   jet (OH)  observed
from the top to the bottom on axis,  from different viewing angles  using the model described
by  \citet{yamazaki04}. From top to bottom: $\theta_o$=0, 1, 1.03, 4  $\theta_j$, where $\theta_{j}=6.1^{\circ}$. The estimated viewing angle for a GRB is
$\theta_{o,grb}$=$\theta_{j}$ (blue dotted line) and for an XRR/XRF is  $\theta_{o,xr}$=$1.03\theta_{k}$ (red dashed line) (see electronic version for color figure).
 The other parameters are:  $\Gamma_{0}=500$, the rest frame frequency $=10\,keV$,
$\epsilon_{b}=0.01$, $\epsilon_{e}=0.1$, z=1, p=2.5, n=10 $ cm^{-1}$.
The straight
line marks the  the time ($20\,ks$) when  we extrapolated the observed
fluxes.}
\label{jetomogeneo}
\end{figure}

Nevertheless, selection effects may weaken these conclusions.  
 We have been assuming that the two classes of events in Fig.6 have the same
mean redshift. This assumption is tentatively supported but not proven true 
by the comparison of the distributions of GRBs and XRFs with known redshift.
 In fact, the mean value of the redshift is compatible  within 1 $\sigma$.  
Our sample of \xxr~ seems to be biased towards high $E_{iso}$ (for a given $E_p$),
allowed by the scattering in the Amati relation. This consideration comes from a direct comparison between
our 3 events with know redshift and the ``Amati'' relation as reported by Ghirlanda et al (2004; see. Fig. \ref{amati}). We have further selected events with an early afterglow
observations. We cannot  exclude that early follow up observations
have been carried out following criteria linked to the property of
the prompt emission (e.g. bright bursts).
These uncertainties on the selection effects do not allow us to draw strong conclusions and, in particular, the QUG cannot be ruled out. 

We now consider the temporal behavior of the X-ray afterglow
lightcurves in the three models
(Figs.~\ref{jetomogeneo}-\ref{jetgaussiano}) and compare it with our
results (\S 4.2 and Figs.~\ref{light_curve} and
~\ref{curve_l_x}).
At early times, the lightcurve
 is remarkably  different as a  function of $\theta_o$ in
the three scenarios.
The sharp edges of the homogeneous jet imply that no light is emitted
along the line of sight for $\theta_o>\theta_j+1/\Gamma$. This gives
the characteristic rising temporal slope as the fireball decelerates
and $1/\Gamma<\theta_o$ (Fig.~\ref{jetomogeneo}). It also implies no
jet breaks in the XRF lightcurve, unless $\theta_o \sim \theta_j$. The UP jet
lightcurves, instead, always have  the temporal evolution of an  on-axis
curve from a homogeneous jet (Fig.~\ref{jetinomogeneo}). An
intermediate behavior is presented by the Gaussian jet (Fig.~\ref{jetgaussiano}): as $\theta_o$
increases, the lightcurve becomes flatter and it eventually recovers
the off-axis behavior for a homogeneous jet.

Unfortunately, our sample is  biased towards  viewing angles close to the jet core/aperture,
where the lightcurve behaviour from then three jet structures becomes very similar.
Thus, a comparison with our current data does not allow us to
discriminate between the models. A  Gaussian jet seen at small angles 
($(\theta_o-\theta_j) \sim (0.1-1) \theta_j$) has been also claimed by
 \citet{granot} et al.  to explain the afterglow of \object{XRF 030723} and \object{XRF 041006}.


Another test that in principle could discriminate between the jet
energy profiles is the ratio between the afterglow X-ray flux and the
prompt $\gamma$-ray flux. This gives a robust estimate of the ratio of
radiation efficiencies of the prompt and afterglow phases, if the
emission is dominated by the line of sight part of the jet
\citep{freedman}.  This is the case for GRBs in all three jet
models. This is also true for \xxr~ in the UP jet model and for the
QUG jets seen close to the jet core.  In those cases, we expect a
similar flux ratio for XRRs/XRFs and GRBs, if the efficiency ratio is
constant with the angular distance from the jet axis.  Instead, the flux
ratio may strongly depend on the viewing angle and on the Lorentz
factor for the OH jet model.  In fact, we find that the two classes
have a mean flux ratio $log(F_{x}/F[40,700])_i$ not compatible at a 
$3\sigma$ level, with a difference of one order of magnitude.
In particular, we may assume that the afterglow
 efficiency is the same in \xxr~ and GRBs: therefore the mechanism
 responsible for the prompt emissio would be more efficient for GRBs.

Finally, the three jet profiles predict a larger
width of the X-ray flux distribution of the XRRs/XRFs, compared to
the GRB one. This is due to the larger $E_{iso}$ distribution expected
for \xxr~(see e.g. Fig.~2 in \citet{yamazaki04}).
 We found that
$\sigma_i=0.45^{+0.14}_{-0.12}$ for GRBs and
$\sigma_i=0.85^{+0.46}_{-0.30}$ for XRRs/XRFs; these values are
compatible within the errors.
Also in this case  we may miss the dim/soft 
events; therefore, the observed XRR/XRF flux distribution can seem  narrower than  the intrinsic one.

\section{Conclusions}

\label{sec:conclusioni}
In this work, we studied the prompt and afterglow emission properties
of XRRs/XRFs, compared to GRB ones. We compiled a sample of 54
XRRs/XRFs and we classified them according to the same hardness
spectral ratio.

We analyzed the spectral parameter distributions of the prompt
emission and we found that the XRR/XRF Band spectral indexes \citep{band}
$\beta_1$ and $\beta_2$ are consistent with GRB ones, while the
peak energy is lower by a factor $\sim4.5$.

We analyzed the light curves of the XRRs/XRFs; we found evidence of breaks both in X-ray and optical afterglow. It is  supported also by the observed different values  of the temporal decay index for  the early and   late afterglow.

We found that the optical and X-ray flux distributions are
consistent for  GRBs and XRRs/XRFs; the ratios   are respectively
of $ \langle HR_{o} \rangle=0.9\pm1.1$ and $ \langle HR_{x}
\rangle=0.9\pm1.2$;

We  discussed our results in the framework of the high-redshift
and the off-axis scenarios.

While the prompt emission spectral
parameters are consistent with a redshifted spectrum, the X-ray  and
optical afterglow properties and the (few) measured redshifts
argue against the interpretation of \xxr~ as high redshift GRBs.

 However, there are also some XRR/XRF events  with an intrinsic energy peak consistent with that of classical GRBs. This suggests  a high redshift nature for them.

 Our analysis of the off-axis scenario favors the OH and UP jet models: both
the X-ray flux ratio between \xxr~ and GBRs and the light curve behavior seem
consistent. 
However,  the prediction of the X-ray flux ratio of the UP jet is parameter-dependent.
The QUG jet  may still be consistent with our results,
if selection effects have reduced the distance between the mean X-ray
fluxes of \xxr~ and GRBs. 

 Finally, our result on the prompt over afterglow flux ratio are
 inconsistent with the simple picture of a constant prompt over
 afterglow efficiency ratio for the UP and QUG jet models.

These conclusions should be further tested, collecting a larger sample
of XRRs/XRFs with known redshift and early afterglow observations, 
allowing one to compare direct luminosity distributions.
The \swift data will be important for this goal, even if this satellite is
sensitive only at energies greater than 15 $keV$, i.e. above the range
of the softest XRFs. Future missions able to select dim/soft
XRRs/XRFs with high sensitivity  can provide  a
fundamental step  to understand the origin of these events.

\paragraph*{Acknowledgments}
The authors are grateful to Enrico  Massaro  for his
suggestions and advice.
We thank Ehud Nakar and Edo Berger for useful discussion and comments as  the anonymous  referee for  very useful suggestions.
Support for this work was provided by NASA though Chandra
Postdoctoral Fellowship grant number PF5-60040
awarded by the Chandra X-ray Center, which is operated by the
smithsonian Astrophysical Observatory for NASA under
contract NASA8-03060.
This work was partially supported by the EU FPS RTN ``Gamma Ray Burst: an enigma and a tool''.

{\scriptsize
\begin{longtable}{lcccccccc}
\caption{ \label{tab1}Spectral parameters of 54 XRRs/XRFs. The  models used   are: BAND=Band function, PL=powerlaw as  N(E)=$KE^{\Gamma}$ , PLE=powerlaw $\times$ exponential cutoff as  N(E)=$KE^{\beta_1}\times exp(-E/E_0)$, PLA= absorbed powerlaw  as N(E)=$KE^{\Gamma}\times exp(-N_H\sigma(E))$. When not given directly, we have derived $\epc =(2+\beta_1)\times \eo$.}\\
\hline\hline
Events & model  & $\beta_1$ & $\beta_2$   &  $E_p (keV)$  &  $\Gamma$ & $N_{H}$ ($10^{22}$cm$^{-2}$) & Reference\\
\hline
\endfirsthead
\caption{(continued)}\\
\hline\hline
Events & model  & $\beta_1$ & $\beta_2$   &  $E_p (keV)$ & $\Gamma$ & $N_{H}$($10^{22}$cm$^{-2}$ & Reference\\
\hline
\endhead
\hline
\endfoot
\object{XRF 971019}   & BAND &  0.98$\pm$0.18           & 3.9$\pm$0.4                 & 19$\pm$1 &  &   &  (1) \\
\object{XRR 971024}               & BAND & -0.4$\pm$1.8           & 2.01$\pm$0.05                  & 5.9$ \pm$ 1.9 &    & & (1) \\
\object{XRR 980128}               & BAND &   1.2$\pm$0.2          & 2.6$\pm$0.7                  & 58$\pm$18 &   &   &(1) \\
\object{XRR 980306}               & BAND &   1.4$\pm$0.2          & 2.5$\pm$0.8                 & 49$\pm$30 &   &    &(1) \\
\object{XRF 981226}[-(180-120)$s$]& PLA  &                        &                        &            & 2.0$\pm$0.4  &  0.018  & (2) \\
\object{XRF 981226}[0-12$s$]      & BAND & 1.3$\pm$0.3            & 2.6$\pm$0.7                   &   61$\pm$15    & & & (2) \\
\object{XRF 981226}[12-36$s$]     & PLA  &                        &                        &               &    2.1$\pm$0.1  &  0.018  &  (2) \\
\object{XRF 981226}[36-51$s$]     & PLA  &                        &                        &                &   2.2$\pm$0.1  &  0.018  &  (2) \\
\object{XRF 981226}[51-84$s$]     & PLA  &                        &                        &               & 2.1$\pm$0.2  &  0.018  &  (2) \\
\object{XRF 990520}               & BAND &  1.3$\pm$0.2           & 3.9$\pm$2.9                 & 26$\pm$3 &  &   & (1) \\
\object{XRF 990526}               & BAND &  1.9$\pm$0.2           &  5                                 & 15$\pm$14    &  & & (1) \\
\object{XRF 990704}[0-6.6$s$]     & BAND & 1.3$^{+0.3}_{-1}$ & 2$^{+4}_{-3}$   &  8$\pm$22 &   & & (3) \\
\object{XRF 990704}[6.6-13.3$s$]  & BAND & 0.92$^{+0.3}_{-0.32}$  &  $2.7^{+0.2}_{-0.2}$      & 8$\pm3$    & &  &  (3) \\
\object{XRF 990704}[13.3-19.5$s$] & BAND & 1.3$^{+0.2}_{-0.3}$ & 2.3$^{+2.8}_{-0.1}$   & 7$\pm16$  &   &  & (3) \\
\object{XRF 990704}[19.5-35.5$s$] & PL   &                        &                              &      & 2.1 $\pm0.1$  &   & (3) \\
\object{XRR 991106}(4)         &      &                        &                               &                      &       & & (5) \\
\object{XRF 000206}               & BAND & 1.6$\pm$0.1            & 5                            & 38$\pm$5  &    & & (1) \\
\object{XRR 000208}               & BAND & 1.4$\pm$0.1            & 3.1$\pm$4.6                  & 87$\pm$36     &  & & (1)\\
\object{XRF 000416}               & BAND & 1.9$\pm$0.9            & 2.5$\pm$0.1                  & 1.6$\pm$6.6        & & & (1)\\
\object{XRF 000615}[0-30$s$]      & PLA  &                        &                         &   &  1.9$^{+0.5}_{-0.3}$ & 21$^{+66}_{-19}$  &  (6)\\
\object{XRF 000615}[30-60$s$]     & PLA  &                        &                         &   &  1.9$\pm$0.3 & 0.027  & (6)\\
\object{XRF 000615}[60-120$s$]    & PLA  &                        &                          &  &  2.2$^{+0.4}_{-0.3}$ & 0.027  &  (6)\\
\object{XRF 010213}               & BAND & 1(frozen)              & 3.0$^{+0.2}_{-0.5}$       & 3.4$\pm0.4$  & & & (7)\\
\object{XRF 010225}               & PLE  & 1.3$\pm0.3$ &                              &  32$^{+27}_{-9}$  & & &(7)\\
\object{XRR 010326B}              & PLE  & 1.1$\pm0.3$ &                              & 52$^{+19}_{-11}$  & & &(7)\\
\object{XRR 010613}               & BAND & 0.95$^{+0.33}_{-0.26}$ & 2.0$\pm0.1$       & 46$^{+18}_{-10}$   & & &(7)\\
\object{XRR 010629}               & PLE  & 1.1$\pm0.1$ &                              & 46$^{+5}_{-4}$ &  & &(7)\\
\object{XRR 010921}               & PLE  & 1.55$^{+0.08}_{-0.26}$ &                              & 89$^{+22}_{-14}$ & & & (7)\\
\object{XRF 011019}               & PLE  & 1.43(frozen)           &                              & 19$^{+18}_{-9}$ &  & &(7)\\
\object{XRR 011030}               & PL   &                        &                              &  &   1.8$\pm0.2$ & & (8)\\
\object{XRR 011103}               & PL   &                        &                              &   &    1.7$^{+0.2}_{-0.3}$ & &(7)\\
\object{XRF 011130}               & PL   &                        &                              & $ <3.9$  &  2.7$\pm0.3$ & &(7)\\
\object{XRR 011211}               & BAND & 1.1$^{+0.2}_{-0.4}$    & 2.1$\pm$0.2              & 18$\pm20$ & & & (9)\\
\object{XRF 011212}               & PL   &                        &                             &  &      2.1$\pm0.2$ & & (7)\\
\object{XRR 020124}               & PLE  & 0.79$^{+0.15}_{-0.14}$ &                              & 87 $^{+19}_{-9}$& & & (7)\\
\object{XRR 020127}               & PLE  & 1.0$\pm0.1$ &                              & 100$^{+50}_{-20}$ &  && (7)\\
\object{XRF 020317}               & PLE  & 0.61$^{+0.61}_{-0.52}$ &                              & 28$^{+13}_{-7}$ & & &(7)\\
\object{XRR 020410}               & PLE  & 1.8                    &                             & 900          & & &(10)\\
\object{XRF 020427}               & BAND/PLA & 1(frozen)          & 2.1$^{+0.2}_{-0.3}$    & 2.8$\pm2.8$ & 2.09$^{+0.23}_{-0.22}$  &0.029 & (11)\\
\object{XRF 020625}               & PLE  & 1.14(frozen)           &                              & 8.5$^{+5.4}_{-2.9}$  & & & (7)\\
\object{XRR 020812}               & PLE  & 1.1$\pm0.3$ &                              & 88$^{+110}_{-30}$  & & &(7)\\
\object{XRR 020819}               & BAND & 0.9$^{+0.2}_{-0.1}$ & 2.0$^{+0.2}_{-0.5}$       & 50$^{+18}_{-13}$   & & &(7)\\
\object{XRF 020903}               & PL   &                        &                              &   $<5$     & 2.6$^{+0.4}_{-0.6}$ & &(7)\\
\object{XRR 021004}               & PLE  & 1.0$\pm0.2$ &                              & 80$^{+53}_{-23}$ & && (7)\\
\object{XRF 021021}               & PLE  & 1.33(frozen)           &                              & 15$^{+14}_{-8}$ & & & (7)\\
\object{XRF 021104}               & PLE  & 1.1$\pm0.5$ &                              & 28$^{+17}_{-8}$ & & &(7)\\
\object{XRR 021112}               & PLE  & 0.9$^{+0.4}_{-0.3}$ &                              & 57$^{+39}_{-21}$  & & &(7)\\
\object{XRR 021211}               & BAND & 0.9$\pm0.1$ & 2.2$^{+0.1}_{-0.3}$       & 46$^{+9}_{-7}$& & & (7)\\
\object{XRR 030115}               & PLE  & 1.3$\pm0.1$ &                              & 83$^{+53}_{-22}$  & & &(7)\\
\object{XRR 030323}                & PL   &                        &                               &  	   & 1.6$\pm0.2$ & & (7)\\
\object{XRR 030324}                & PLE  & 1.5$^{+0.1}_{-0.2}$ &                              & 150$^{+630}_{-70}$& & &(7)\\
\object{XRR 030329}                 & BAND & 1.26$^{+0.01}_{-0.02}$ & 2.3$^{+0.1}_{-0.1}$       & 68$\pm2$ & & & (7)\\
\object{XRF 030416}                  & PL   &                        &                              &  $ <3.8$   & 2.3$\pm0.1$ & & (7)\\
\object{XRR 030418}                 & PLE  & 1.5$\pm0.1$ &                              & 46$^{+32}_{-13}$  & & &(7)\\
\object{XRF 030429}                  & PLE  & 1.1$\pm0.2$ &                              & 35$^{+12}_{-8}$& & & (7)\\
\object{XRF 030528}                & BAND & 1.3$^{+0.2}_{-0.1}$ & 2.7$^{+0.3}_{-1}$       & 32$\pm5$ & & & (7)\\
\object{XRF  030723}                  & PL   &                        &                              &   $<8.9$    & 1.9$\pm0.2$ & & (7)\\
\object{XRR 030725}                 & PLE  & 1.51$^{+0.04}_{-0.04}$ &                              & 100$^{+20}_{-10}$  & & &(7)\\
\object{XRR 030821}                 & PLE  & 0.9$\pm0.1$ &                             & 84$^{+15}_{-11}$  & && (7)\\
\object{XRF 030823}                  & PLE  & 1.3$\pm0.2$ &                              & 27$^{+8}_{-5}$  & & &(7)\\
\object{XRF 030824}                     & PL   &                        &                              &  $ <8.7$& 2.1$\pm0.1$&0.5  & (7)\\
\object{XRF 031109B}                 & PLE  & 0.5 &                              & 29 &  & &(11)\\
\object{XRR 031220}                  & PLE  & 1 &                             & 49& & &(11)\\

\end{longtable}
(1){\footnotesize  :\citet{kip03}, (2): \citet{fro00}, (3):\citet{fero01}, (4): possible Tipe I X-ray  Burster, (5): \citet{gand99}, (6): \citet{Miao04}, (7):  \citet{Sakamoto}, (8): \citet{galli}, (9): \citet{piro04},  (10): \citet{nic04},  (11): \citet{ama04},
(12): http:space.mit=http:space.mit.edu/HETE/BURST}
}

{\footnotesize
\begin{longtable}{lccccccc}
\caption{\label{tab2} General  properties  of 54 events XRRs/XRFs: Time UT of the burst trigger, instrument of the observation,  fluences in  $10^{-7}erg\,cm^{-2}$ in the energy range 30-400 \kev (S[30,400]), 2-30 \kev (S[2,30]), 40-700  \kev (S[40,700]) and 2-10\kev (S[2,10])}\\
\hline\hline
Events&Time UT & Instruments  & $S[30,400]$ & $S[2,30]$ & $S[40,700]$ & $S[2,10]$ & Ref  \\
\hline
\endfirsthead
\caption{(continued)}\\
\hline\hline
Events   & Time UT & Instruments  & $S[30,400]$ & $S[2,30]$ & $S[40,700]$ & $S[2,10]$ & Ref \\
\hline
\endhead
\hline
\endfoot
\object{XRF 971019}   &         & WFC/BATSE off-line    &             &           &             &         &    (1)\\
\object{XRR 971024}   &         & WFC/BATSE off-line    &             &           &             &         &    (1)\\
\object{XRR 980128}   &         & WFC/BATSE off-line    &             &           &             &         &    (1) \\
\object{XRR 980306}   &         & WFC/BATSE off-line    &             &           &             &         &     (1) \\
\object{XRF 981226}   & 26.41  & GRBM,WFC              &     &   & 4$\pm1$    &  5.7$\pm1.0$ & (2) \\
\object{XRF 990520}   & 20.09  & WFC/BATSE off-line    &     &   &  8$\pm3$(12) &               & (1) \\
\object{XRF 990526}   &        & WFC/BATSE off-line    &             &           &             &         &   (1) \\
\object{XRF 990704}   & 4.73   & GRBM, WFC             &     &   &  10$\pm1$  & 15.0$\pm$0.8  & (3)  \\
\object{XRR 991106}   &  6.45  & GRBM,WFC              &             &           &   $<1.2$          &         &    (4) \\
\object{XRF 000206}   &        & WFC/BATSE off-line    &             &           &             &         &     (1) \\
\object{XRR 000208}   &        & WFC/BATSE off-line    &             &           &             &         &     (1) \\
\object{XRF 000416}   &        & WFC/BATSE off-line    &     &   & 2.9$\pm$0.3(12)             &         & (1) \\
\object{XRF 000615}   &        & GRBM,WFC              &     &   & 9.8$\pm$0.9 &  17$\pm1$ & (5) \\
\object{XRF 010213}   & 13.53  & FREGATE,WXM           & $0.69^{+0.58}_{-0.32}$ &  $7.9\pm0.3$  & & & (6)\\
\object{XRF 010225}   &        & FREGATE,WXM                      & $2.4^{+1.7}_{-0.9}$ &  $3.5\pm0.4$  & & & (6)\\
\object{XRR 010326B}  & 26.36  & FREGATE,WXM           & $3.2^{+0.9}_{-0.8}$ &  $2.4\pm0.3$  & & & (6)\\
\object{XRR 010613}   & 13.32  & FREGATE,WXM            & $228\pm13$ &  $102\pm7$  & & & (6)\\
\object{XRR 010629}   & 29.52  & FREGATE,WXM           & $29\pm3$ &  $25\pm2$  & & & (6)\\
\object{XRR 010921}   & 21.22  & FREGATE,WXM           & $113^{+9}_{-8}$ &  $72\pm3$  & & & (6)\\

\object{XRF 011019}   & 19.36  & FREGATE,WXM           & $1.1^{+1.4}_{-0.7}$ &  $3.0\pm0.6$  & & & (6)\\
\object{XRR 011030}   & 30.27  & WFC                   &    &    &   &  1.2(13)  & (7) \\
\object{XRR 011103}   &   & FREGATE,WXM         & $6^{+9}_{-3}$ &  $3.3^{+0.8}_{-0.7}$  & & & (6)\\
\object{XRF 011130}   & 30.26  & FREGATE,WXM           & $0.98^{+1.17}_{-0.62}$ &  $5.9\pm0.1$  & & & (6)\\
\object{XRR 011211}    & 11.88  & GRBM, WFC             &     &   &  37$\pm4$  & 11$\pm$1  & (8)  \\
\object{XRF 011212}    & 12.17  & FREGATE, WXM           & $3.4^{+2.5}_{-1.7}$ &  $4.2\pm0.6$  & & & (6)\\
\object{XRR 020124}     & 24.45  & FREGATE,WXM           & $61^{+9}_{-8}$ &  $20\pm1$  & & & (6)\\
\object{XRR 020127}     & 27.87  & FREGATE,WXM           & $21^{+5}_{-4}$ &  $6.7\pm0.5$  & & & (6)\\
\object{XRF 020317}     & 17.76  & FREGATE,WXM           & $1.3^{+0.9}_{-0.6}$ &  $2.2\pm0.4$  & & & (6)\\
\object{XRR 020410}     & 30.27  & WFC, KONUS(WIND)                   &    &    & $\sim 290$  &  $>47$  & (9)\\
\object{XRF 020427}      & 27.18  & WFC                  &   &   & $ <2.9$ &  37$\pm0.3$  & (10) \\
\object{XRF 020625}      & 25.48  & FREGATE,WXM          & $0.12^{+0.35}_{-0.11}$ &  $2.4^{+0.6}_{-0.5}$  & & & (6)\\
\object{XRR 020812}      & 12.45  & FREGATE,WXM           & $19^{+8}_{-6}$ &  $7.9\pm1.1$  & & & (6)\\
\object{XRR 020819}      & 19.62  & FREGATE,WXM,SXC           & $63^{+8}_{-9}$ &  $25\pm1$  & & & (6)\\
\object{XRF 020903}      & 3.42 & FREGATE,WXM,SXC        & $0.16^{+0.44}_{-0.14}$ &  $0.83^{+0.28}_{-0.24}$  & & & (6)\\
\object{XRR 021004}      & 4.50  & FREGATE,WXM,SXC           & $18^{+7}_{-5}$ &  $7.7\pm0.7$  & & & (6)\\
\object{XRF 021021}      & 21.78  & FREGATE,WXM           & $0.62^{+1.07}_{-0.49}$ &  $2.5\pm0.6$  & & & (6)\\
\object{XRF 021104}      & 4.29  & FREGATE,WXM           & $6.1^{+4.4}_{-2.7}$ &  $10\pm2$  & & & (6)\\
\object{XRR 021112}      & 12.15  & FREGATE,WXM           & $2.1^{+1.1}_{-0.9}$ &  $1.3\pm0.3$  & & & (6)\\
\object{XRR 021211}       & 11.47  & FREGATE,WXM,SXC           & $23.71^{+2.03}_{-2.01}$ &  $11.6\pm0.3$  & & & (6)\\
\object{XRR 030115}       & 15.14  & FREGATE,WXM,SXC           & $15^{+4}_{-3}$ &  $7.9\pm0.6$  & & & (6)\\
\object{XRR 030323}       & 23.92  & FREGATE,WXM,SXC           & $8.9^{+3.8}_{-3.5}$ &  $3.4^{+1.3}_{-1.2}$  & & & (6)\\
\object{XRR 030324}       & 24.13  & FREGATE,WXM           & $13\pm3$ &  $5.5^{+0.4}_{-0.5}$  & & & (6)\\
\object{XRR 030329}       & 29.48  & FREGATE,WXM,SXC           & $1076^{+13}_{-14}$ &  $553\pm3$  & & & (6)\\
\object{XRF 030416}        & 16.46  & FREGATE off-line,WXM         & $3.7^{+1.9}_{-1.4}$ &  $9.0\pm0.9$  & & & (6)\\
\object{XRR 030418}        & 18.42  & FREGATE,WXM           & $17^{+7}_{-5}$ &  $17\pm1$  & & & (6)\\
\object{XRF 030429}        & 29.41  & FREGATE,WXM,SXC           & $3.8^{+1.4}_{-1.2}$ &  $4.7\pm0.5$  & & & (6)\\
\object{XRF 030528}        & 28.55 & FREGATE,WXM           & $ 56\pm7$ &  $63\pm3$  & & & (6)\\
\object{XRF 030723}         & 23.27  & FREGATE,WXM,SXC           & $0.38^{+5.56}_{-0.33}$ &  $2.8\pm0.5$  & & & (6)\\
\object{XRR 030725}          & 25.49  & FREGATE,WXM           & $167\pm10$ &  $94\pm2$  & & & (6)\\
\object{XRR 030821}          & 21.23  & FREGATE,WXM           & $28^{+3}_{-2}$ &  $10\pm0.6$  & & & (6)\\
\object{XRF 030823}          & 23.37  & FREGATE,WXM,SXC           & $13\pm4$ &  $23\pm2$  & & & (6)\\
\object{XRF 030824}          & 24.70  & FREGATE,WXM           & $5.8^{+2.4}_{-1.9}$ &  $8.9\pm1.1$  & & & (6)\\
\object{XRF 031109B}         & 19.07 & FREGATE,WXM     & & & & & (11) \\
\object{XRR 031220}          & 20.15 & FREGATE,WXM,SXC  & & & & & (11)\\

\end{longtable}

(1){\small: \citet{kip03}, (2): \citet{fro00}, (3): \citet{fero01}, (4): \citet{gand99}, (5): \citet{Miao04}, (6): \citet{Sakamoto}, (7): \citet{galli}, (8): \citet{piro04},  (9): \citet{nic04},  (10): \citet{ama04},
(11): http:space.mit=http:space.mit.edu/HETE/BURST, (12):fluence between $50-300$$keV$, (13):fluence between $2-28$$keV$ }

}

\newpage

\begin{table*}[t]
\caption{ Hardness Ratio $ H_h=S[2,30]/S[30,400]$ and  $ H_s=S[2,10]/S[40,700]$ for 54 XRRs/XRFs. }
\label{tab3}
\begin{tabular}{lcclcc}
\hline
Events   & $H_s$  & $H_h$   & Events   & $H_s$  & $H_h$ \\
\hline

\object{XRF 971019}  &  2.1$^{+1.4}_{-0.9}$     &    3.0 $^{+1.2}_{-2.1}$      &  \object{XRR 020410}    & 0.27  $^{(7)}$      &  0.62       \\
\object{XRR 971024}    & 0.1$^{+0.5}_{-0.1}$       & 0.5$^{+0.8}_{-0.1}$            &        \object{XRF 020427}      &    8.4 $^{(8)}$    &    1.3$\pm0.1$     \\
\object{XRR 980128}    &    0.2$^{+0.2}_{-0.1}$    &    0.6$^{+0.3}_{-0.1}$        &        \object{XRF 020625}     &   36$^{+900}_{-30}$      &    21 $^{(5)}$    \\
\object{XRR 980306}     &   0.3$^{+0.2}_{-0.1}$     &    0.7$^{+0.4}_{-0.2}$       &       \object{XRR 020812}   &    0.2$^{+0.1}_{-0.1}$     &  0.41  $^{(5)}$     \\
\object{XRF 981226} &     1.4$\pm$0.4 $^{(1)}$   &  1.2$^{+0.5}_{-0.1}$     &                 \object{XRR 020819}    &   0.1$^{+0.1}_{-0.03}$      & 0.4 $^{(5)}$ \\
\object{XRF 990520} & 1.1$^{+0.9}_{-0.4}$       &    1.8$^{+0.9}_{-0.5}$               &      \object{XRF 020903}   &   5.0$^{+11}_{-4}$      &   7.3 $^{(5)}$     \\
\object{XRF 990526} &   1.2$^{+0.8}_{-0.3}$     &    1.8 $^{+0.9}_{-0.5}$               &    \object{XRR 021004}   & 0.2 $^{+0.4}_{-0.1}$       &   0.43 $^{(5)}$     \\
\object{XRF 990704}  &    1.5$\pm0.2$  $^{(2)}$   &   1.2$^{+0.2}_{-1.0}$    &               \object{XRF 021021}   &  3.5$^{+33}_{-2.1}$       &  4.0 $^{(5)}$   \\
\object{XRR 991106} &   $>0.8$  $^{(3)}$   &             &         \object{XRF 021104}   & 0.97$^{+20}_{-0.5}$        &    1.7 $^{(5)}$    \\
\object{XRF 000206} &    0.7$^{+0.3}_{-0.2}$     &  1.2$^{+0.4}_{-0.2}$          &             \object{XRR 021112}   &  0.2 $^{+1.1}_{-0.1}$      &  0.61 $^{(5)}$      \\
\object{XRR 000208} &   0.2$^{+0.2}_{-0.1}$     &    0.6 $^{+0.3}_{-0.2}$       &              \object{XRR 021211}   & 0.2 $\pm0.1$        &  0.49 $^{(5)}$      \\
\object{XRF 000416} & 3$^{+0.2}_{-1.1}$       &    3.7$^{+0.5}_{-2.3}$        &              \object{XRR 030115}   & 0.2$^{+0.2}_{-0.1}$        & 0.52   $^{(5)}$     \\
\object{XRF 000615}  & 1.7$\pm$0.2 $^{(4)}$      & $ >1.48$ &               \object{XRR 030323}   &    0.1$\pm0.1$     &  0.38  $^{(5)}$   \\
\object{XRF 010213}  &  3.3$\pm0.7$      &     11 $^{(5)}$    &               \object{XRR 030324}   & 0.2       & 0.52  $^{(5)}$      \\
\object{XRF 010225}  & 0.8$^{+2.7}_{-0.4}$     &    1.5 $^{(5)}$   &                  \object{XRR 030329}   &  0.2$^{+0.3}_{-0.1}$       & 0.51    $^{(5)}$    \\
\object{XRR 010326B}   & 0.3$^{+0.6}_{-0.1}$      &  0.75 $^{(5)}$     &                \object{XRF 030416}   &  1.7$^{+0.8}_{-0.4}$       &  2.4  $^{(5)}$     \\
\object{XRR 010613}    &    0.1$\pm0.1$   &   0.45 $^{(5)}$   &                 \object{XRR 030418}    &  0.5$^{+0.3}_{-0.1}$       & 0.99   $^{(5)}$     \\
\object{XRR 010629}    &   0.5 $^{+0.4}_{-0.3}$    &  0.9  $^{(5)}$              &      \object{XRF 030429}   &  0.7 $^{+1}_{-0.3}$      & 1.3  $^{(5)}$      \\
\object{XRR 010921}     &  0.3$^{+0.4}_{-0.1}$      &  0.64  $^{(5)}$        &            \object{XRF 030528}   &  0.6$\pm0.1$       & 1.1  $^{(5)}$      \\
\object{XRF 011019}     &    2.1$^{+4.6}_{-1.1}$    & 2.8  $^{(5)}$         &           \object{XRF 030723}   &  0.4 $^{+0.4}_{-0.2}$      &  7.5  $^{(5)}$     \\
\object{XRR 011030}     &  0.3 $^{+0.3}_{-0.1}$     &     0.7$^{+0.4}_{-0.2}$  &                      \object{XRR 030725}   &  0.3$\pm0.1$       &   0.56  $^{(5)}$    \\
\object{XRR 011103}   &    0.2$^{+0.2}_{-0.1}$    &  0.53   $^{(5)}$           &         \object{XRR 030821}   &   0.1$\pm0.1$      &  0.36  $^{(5)}$     \\
\object{XRF 011130}   &  5.3$^{+12}_{-3}$       &   6    $^{(5)}$         &        \object{XRF 030823}   & 1.1 $^{+1.4}_{-0.1}$       &  1.8    $^{(5)}$    \\
\object{XRR 011211}   &  0.3 $^{(6)}$     &    0.9$\pm0.4$   &                     \object{XRF 030824}   &  0.9 $^{+0.2}_{-0.4}$      & 1.5  $^{(5)}$      \\
\object{XRF 011212} &   0.7$^{+0.9}_{-0.3}$     & 1.3  $^{(5)}$             &          \object{XRF 031109B}    &  0.73      &   1.3     \\
\object{XRR 020124}   &  0.1$\pm0.1$      &  0.32    $^{(5)}$       &           \object{XRR 031220}   &  0.32      &  0.77      \\
\object{XRR 020127}    & 0.1$\pm0.1$        &  0.33  $^{(5)}$        &    & & \\
\object{XRF 020317} &  0.8$^{+20}_{-0.5}$          &   1.7    $^{(5)}$    &    &  &  \\
\hline
\multicolumn{6}{c} { }
\end{tabular}

(1){\small=\citet{fro00}, (2)=\citet{fero01}, (3)=\citet{gand99}, (4)=\citet{Miao04}, (5)=\citet{Sakamoto}, (6)=\citet{piro04}, (7)=\citet{nic04}, (8)=\citet{ama04}} \\

\end{table*}
\newpage

\begin{table*}
\caption{Mean values  and standard deviation  of  spectral parameters $\beta_1$, $\beta_2$ and $log\epc$
 for  the  parent distribution   of XRRs/XRFs and GRBs classes, according to the likelihood method. In the last row is  the $log\epc$ instrinsic mean value for the subsample of events used  for the X-ray afterglow flux analysis in \S~\ref{sec:results}.}
 \label{tab10}
  \begin{tabular}{|rcc|ccc|}
   \hline
     CLASS(number)&$\langle\beta_1\rangle_i$&$\langle\sigma\rangle_i$ &CLASS(number)& $\langle\beta_2\rangle_i$ &$\langle\sigma\rangle_i$\\
   \hline
     XRRs+XRFs (38) & $-(1.22^{+0.12}_{-0.09})$ & 0.21$^{+0.12}_{-0.09}$ & XRRs+XRFs (25) & -(1.74$\pm$0.42) & 1.17$^{+0.42}_{-0.28}$ \\
     GRB (31)      & $-(0.99^{+0.08}_{-0.10}$) & 0.28$^{+0.06}_{-0.05}$  & GRB (19)      & $-(2.31^{+0.20}_{-0.16})$  & 0.46$^{+0.20}_{-0.15}$   \\
   \hline
     CLASS (number)     & $\langle Log E_{peak}\rangle_i$ & $\langle\sigma\rangle_i$ &CLASS(number)& $\langle\tilde{Log E}_{p}\rangle_i$ &$\langle\sigma\rangle_i$\\
   \hline
     XRRs+XRFs (42) & 1.55$\pm$0.1   &  0.32$\pm$0.10    &XRRs+XRFs(10)  &$2.13\pm0.11$ & $0.17^{+0.16}_{-0.12}$ \\
     GRB (30)      & $2.21^{+0.06}_{-0.07}$   & 0.25$^{+0.08}_{-0.05}$ &GRBs(12) &$2.74^{+0.14}_{-0.16}$ &$0.32^{+0.16}_{-0.09}$  \\
  \\
   \hline
     SUBCLASS (number)     & $\langle Log E_{peak}\rangle_i$ &   & &   & \\
   \hline
     XRRs+XRFs (7) & 1.83$\pm$0.06   &  &  &  &  \\
     GRB (14)      & $2.32\pm0.10$   &  &  & & \\
   \hline

 \end{tabular}
\end{table*}

\newpage

\begin{table*}

\caption{Intrinsic Peak Energy, Isotropic Energy [1,1000 keV] and redshift values  or constraints  for 14  XRRs/XRFs: the TYPE column indicates  if the measure of the redshift  is obtained from  Host Galaxy spectroscopy (HG), or from  Optical afterglow spectroscopy (OT). We report also the pseudo-redshift for 6 possibly high redshift GRBs(16). }
\label{tab7}
\begin{tabular}{lcccccc}
\hline
Burst &$ \tilde{E}_{p}$(keV) & $ E_{iso}$ ($10^{52}$ erg) & $z$&TYPE & Burst& pseudo-redshift\\
\hline
\object{XRR 010921}&$129\pm32$&0.68 &0.45&HG (1)  &   \object{XRF 990520}  &$0.6\pm1.0$           \\
\object{XRR 011030}&& &$<3.5$&HG (2) &  \object{XRF 990704}  & $0.2\pm0.1$                  \\
\object{XRR 011211}&$56\pm63$&1.7 & 2.14&OT (3)  &  \object{XRF 011019}  &  $0.6\pm0.6$                    \\
\object{XRR 020124}&$365\pm80$&3.7 &3.2&OT (4)  &    \object{XRF 021104}  & $1.2\pm1.1$ (8)                    \\
\object{XRF 020427}&$<9$&$<0.08$  &$<2.3$&HG (2)&   \object{XRR 021112}  &   $4.6\pm4.3$                    \\
\object{XRF 020903}&$<6$&$4.8\times10^{-5}$  &0.25&HG (5)  &    \object{XRF 030823}  &    $0.8\pm0.7$ (8)                   \\
\object{XRR 021004}&$266\pm176$&0.9  &2.33&OT (6)  &                       &    \\
\object{XRR 021211}&$93\pm18$&0.5  &1.01&HG (7)   &                       &    \\
\object{XRR 030115}&$129\pm32$&0.68  &2.2&HG (8)  &                          &     \\
\object{XRR 030329}&$80\pm3$&1.03  &0.17&OT (9)   &                   &   \\
\object{XRR 030323}&& &$3.37$&HG (10) &                   &            \\
\object{XRR 030324}&$<555$&  &$<2.7$&HG (11) &                    &          \\
\object{XRR 030429}&$128\pm44$&0.4  &2.65&OT (12)  &                    &            \\
\object{XRF 030528}&$57\pm9$&1  &0.78&HG (13) &                         &    \\
\object{XRF 030723}&$<28$&$<0.5$  &$<2.1$&OT (14)  &                      &        \\
\object{XRR 031220}&$142\pm15$&0.05  &$1.90\pm0.30$&OT (15) &               &               \\

\hline
\multicolumn{7}{c} { }
\end{tabular}

(1){\small=\citet{bloom1}, (2)=\citet{blooma}, (3)=\citet{fruch1}, (4)=\citet{hior}, (5)=\citet{lev2}, (6)=\citet{cor}, (7)=\citet{dellavalle},(8)=\citet{pelangeon}, (9)=\citet{bloomB},(10)=\citet{vreeswijk},  (11)=\citet{ny4}, (12)=\citet{weid03}, (13)=\citet{rau05}, (14)=\citet{fynbo}, (15)=\citet{mel} (16)=http://www.ast.obs-mip.fr/grb/pz}\\
\end{table*}

{\scriptsize
\begin{longtable}{lcccccc}
\caption{ \label{tab4} Afterglow properties of  54 XRRs/XRFS:  X-ray ToO observations (AX), time in days of the start of  ToO and  satellite which performed the observation,  optical afterglow detection (OT), time   of the OT detection and observed  magnitude,  Radio afterglow detection (RT) and  host galaxy detection}\\
\hline\hline
Events & A X & ToO Date[$d$] and istrument &OT& Date[$d$] and $m_R$[$m$] OT & Radio T & Host galaxy  \\
\hline
\endfirsthead
\caption{(continued)}\\
\hline\hline
Events  & A X & ToO Date[$d$] and istrument &OT& Date[$d$] and $m_R$[$m$] OT & Radio T & Host galaxy  \\
\hline
\endhead
\hline
\endfoot
\object{XRF 971019}   &  -       & -                  &     -      &      -        &     -    &    -     \\
\object{XRR 971024}   &   -      & -                  &    -       &     -         &     -    &    -     \\
\object{XRR 980128}   &   -      & -                  &     -      &     -         &     -    &    -     \\
\object{XRR 980306}   &   -      & -                  &      -     &      -        &     -    &    -     \\
\object{XRF 981226}   & Y$^{1}$        &   0.3, 7.2 ($\sax$)  &  N$^{2}$       & 0.40  $m >23$ & Y$^{3}$        & Y?$^{4}$  \\
\object{XRF 990520}   & -        & -                   &  N$^{5}$         & 0.74 $m>22.5 $ &  N  $^{6}$     &    -      \\
\object{XRF 990526}   &  -       & -                  &     -      &      -        &     -    &    -      \\
\object{XRF 990704}   & Y $^{7}$       & 0.3,7.1 ($\sax$)    &  N $^{8}$        & 0.19 $ m>22.5$  &  N $^{9}$      &  -       \\
\object{XRR 991106}   & Y? $^{10}$       & 0.33  ($\sax$)    &  N $^{11}$        & 0.42  $m>22$  &  N $^{12}$      &  -        \\
\object{XRF 000206}   &  -       & -                  &     -      &      -        &     -    &    -     \\
\object{XRR 000208}   &  -       & -                  &     -      &      -        &     -    &    -     \\
\object{XRF 000416}   &  -       & -                  &     N$^{13}$      & 2.64 $m>20.7$        &     -    &    -     \\
\object{XRF 000615}   & Y$^{14}$        & 0.42   ($\sax$)    &  N$^{15}$         & 0.18  $m>21.5$  &  -       &  -      \\
\object{XRF 010213}   & -        & -                  &  N $^{16}$        & 1.98  $m>22$  &  N  $^{17}$     &    -   \\
\object{XRF 010225}    &   -      & -                  &      -     &      -        &     -    &    -     \\
\object{XRR 010326B}   &   -      & -                  &      N$^{18}$     & 0.5 $m>21$             &     -    &    -     \\

\object{XRR 010613}     &   -      & -                  &      -     &      -        &     -    &    -     \\
\object{XRR 010629}     &   -      & -                  &      N $^{19}$    & 0.49  $m>20.5$             &     -    &    -     \\
\object{XRR 010921}      &   -      & -                  &      Y$^{20}$     & 1.05 $m=19.9\pm0.2$             &     -    &    Y    \\
             &   -      & -                  &          & 0.92  $m>20.5$  $^{21}$           &     -    &    -     \\
\object{XRF 011019}       &   -      & -                  &      N$^{22}$     & 1.07  $m>22.5$             &     -    &    -     \\
\object{XRR 011030}       & Y $^{23}$       & 11, 31.2    ($\chandra$)    &  N$^{24}$         & 0.30  $m>21$  &  Y   $^{25}$    &  Y?$^{26}$  \\
\object{XRR 011103}       &   -      & -                  &      -     &      -        &     -    &    -     \\
\object{XRF 011130}       & Y $^{27}$       & 10, 83    ($\chandra$)    &  Y? $^{28}$        & 6.89  $m=23.0\pm0.1$  &  Y? $^{29}$       &  Y  $^{30}$\\
\object{XRR 011211}       & Y $^{31}$       & 0.5    ($\xmm$)    &  Y $^{32}$        & 0.46  $m=20.41\pm0.04$   &  -     &  Y  $^{33}$\\
\object{XRF 011212}       &   -      & -                  &      Y?$^{34}$     & 2.00  $m=23.8\pm0.2$             &     -    &    -     \\
\object{XRR 020124}       &   -      & -                  &      Y $^{35}$    & 1.51 $m=23.84\pm0.17$              &     -    &    -     \\
\object{XRR 020127}       & Y   $^{36}$& 4.1, 14.6     ($\chandra$)    &  N  $^{37}$        & 0.19  $m>19.5$  &  Y  $^{38}$      &  Y$^{39}$   \\
\object{XRF 020317}       &   -      & -            &      Y?  $^{40}$    & 0.76  $m=19.6\pm0.1$             &     -    &    -     \\
\object{XRR 020410}       & Y  $^{41}$       & 0.83, 2.3     ($\sax$)    &  Y  $^{41}$         & 0.26  $m=21.0\pm0.5$  &  N  $^{42}$&  -           \\
\object{XRF 020427}       & Y $^{43}$        & 0.46,2.5 ($\sax$)    &  -        & -  &  Y? $^{44}$       &  Y? $^{45}$            \\
             &          & 9.2,17.2($\chandra$)  &         &         &    &     \\
\object{XRF  020625}    &   -      & -                  &      N $^{46}$     & 0.44   $m>18.2$             &     -    &    -     \\
\object{XRR 020812}    &   -      & -                  &      N  $^{47}$    & 0.13  $m>19$             &     -    &    -     \\
\object{XRR 020819}    &   -      & -                  &      N  $^{48}$    & 0.72  $m>21.7$             &     Y  $^{49}$   &    Y? $^{50}$ \\

\object{XRF 020903}    &   -      & -                  &      Y $^{51}$     & 26.5   $m=18.6\pm0.2$&    Y? $^{52}$   &    Y  $^{53}$   \\
          &         &                   &            & 1.00  $m>19.8$ $^{54}$  &     & \\
\object{XRR 021004}      & Y$^{55}$        & 0.85, 52.3($\chandra$)    &  Y$^{56}$& 0.48  $m=18.46\pm0.05$   &  Y$^{57}$&  Y$^{58}$     \\
\object{XRF 021021}    &  -       & -                  &     -      &      -        &     -    &    -     \\
\object{XRF 021104}     &   -      & -             & N $^{59}$    & 0.12  $m>21$             &     -    &    -     \\
\object{XRR 021112}   &   -      & -                  &      N $^{60}$    & 0.08  $m>21$             &     N$^{61}$    &    -     \\
\object{XRR 021211}    &   -      & -                  &      Y $^{62}$    & 0.85  $m=23.20\pm0.18$             & N $^{63}$   & Y$^{64}$      \\
\object{XRR 030115}    &   -      & -                  &      N$^{65}$     & 0.48  $m>21$             &Y $^{66}$  &Y$^{67}$     \\
\object{XRR 030323}   &   -      & -                  &   Y $^{68}$   & 0.35  $m=18.83\pm0.09$             & -   &  Y$^{69}$      \\
\object{XRR 030324}    &   -      & -                  &      N$^{70}$     & 0.39  $m>21.4$             &     N$^{71}$    &Y?$^{72}$     \\
\object{XRR 030329}     & Y$^{73}$ & 0.2,1.3    ($\rxte$)    &  Y $^{74}$        & 0.37  $m=14.55\pm0.03$  &  Y $^{75}$ &  Y$^{76}$       \\
             &          & 37, 61, 258($\xmm$)  &         &         &    &     \\
\object{XRF 030416}  &   -      & -                  &      N$^{77}$     & 1.45  $m>15$             &     -    &    -     \\
\object{XRR 030418} &   -      & -                  &      Y$^{78}$     & 6.58  $m=24.9\pm0.4$             &    -     &   -      \\
\object{XRF 030429}  &   -      & -                  &      Y$^{79}$     & 0.37  $m=20.20\pm0.15$             & N$^{80}$         &    Y$^{81}$  \\
\object{XRF 030528}  & Y$^{82}$        & 6, 12    ($\chandra$)    &  N$^{83}$       & 0.002  $m>18.7$  &  -       &  Y $^{84}$       \\
\object{XRF 030723}  & Y $^{85}$       & 2.1, 12.7     ($\chandra$)    &  Y $^{86}$        & 0.86  $m=21.9\pm0.2$  &  N$^{87}$      &  -      \\
\object{XRR 030725}    &   -      & -                  &      Y$^{88}$     & 3.89  $m=21.2\pm0.2$             &   -    &  -    \\
\object{XRR 030821}    &  -       & -                  &     -      &      -        &     -    &    -     \\
\object{XRF 030823}   &   -      & -                  &      N$^{89}$     & 0.78  $m>22.5$             &     -    &    -     \\
\object{XRF 030824}    &   -      & -                  &      N$^{90}$     & 1.72  $m>22.5$             &     -    &    -     \\
\object{XRR 031109B}   &  -       & -                  &     -      &      -        &     -    &    -     \\
\object{XRR 031220} & Y$^{91}$        & 5.6, 28.5   ($\chandra$)    &  N$^{92}$         & 0.23  $m>21$  &  -       &  -          \\
\end{longtable}

(1){\scriptsize : \citet{fro00},   (2): \citet{lind},   (3): \citet{frail}, (4): \citet{holland}, (5): \citet{ct},  (6): \citet{frail2},
(7): \citet{fero01}, (8): \citet{jens}, (9): \citet{rol}, (10): \citet{anto},  (11): \citet{jensb}, (12): \citet{frailB},   (13): \citet{price},  (14): \citet{nic01},  (15): \citet{stan00}   (16): \citet{zhu},  (17): \citet{ber01A},  (18): \citet{price4},  (19): \citet{and},   (20): \citet{priceb, lamb01, bloom1},  (21): \citet{ser}, (22): \citet{kom01},    (23): \citet{heise},  (24): \citet{moh}, (25): \citet{harr},
(26): \citet{blooma},  (27): \citet{ricker, but},  (28): \citet{gar}, (29): \citet{ber01B},   (30): \citet{gar}, (31): \citet{piro04}, (32): \citet{ses},    (33): \citet{bur},   (34): \citet{dul}, (35): \citet{gor}, (36): \citet{fox},  (37): \citet{lamb02},  (38): \citet{rol2}, (39): \citet{fox2},
(40): \citet{tom}, (41): \citet{nic04}, (42): \citet{frail3},  (43): \citet{ama04}, (44): \citet{wi},  (45): \citet{ct2,blooma}, (46): \citet{burn},
(47): \citet{kaw},   (48): \citet{pic02}, (49): \citet{hen},   (50): \citet{lev},  (51): \citet{ste},  (52): \citet{ber02A},  (53): \citet{lev2}, (54): \citet{price2},  (55): \citet{rsak2, rsak1}, (56): \citet{sat},  (57): \citet{frail02},  (58):  \citet{dostoj2}, (59): \citet{foxP2},
(60): \citet{sch},              (61): \citet{frailBer},  (62): \citet{mac02}, (63): \citet{ber2F},   (64): \citet{fruch2}, (65): \citet{ny},
(66): \citet{ber2F3},    (67): \citet{mas03},   (68): \citet{woo}, (69):  \citet{vreeswijk},  (70):  \citet{ric2} ,  (71): \citet{ber2F34},  (72): \citet{ny2},
(73): \citet{marshall,  marshall2, tiego2, schartel, tiego3}, (74): \citet{bar},   (75): \citet{ber223},     (76): \citet{bloomB},  (77): \citet{lip},    (78): \citet{dul3},
(79): \citet{rum}, (80): \citet{ber2F343},  (81): \citet{kam},  (82): \citet{but3}, (83): \citet{ay},(84): \citet{rau04}, (85): \citet{but3c, but3b},   (86): \citet{fynbo},
(87): \citet{fynbo3},  (88): \citet{vin}, (89): \citet{foxP}, (90):\citet{foxP1} , (91): \citet{gen04},  (92): \citet{fox3}}

}

{\footnotesize

\begin{longtable}{lcccccc}
\caption{ \label{tab5}The   X-ray $(F_{x})$   and   optical $(F_{o})$ flux of the afterglow  at 40 $ks$  and their ratio $ (F_{o/x})$; $N_H$ is the hydrogen column density, $\delta_o$ is the temporal index decay of the optical afterglow  and  $\delta_x$ is the temporal decay index of the  X-ray afterglow }\\
\hline\hline
Events  & $N_H^{(1)}$ & $F_{o}^{(2)}$   & $\delta_o$ & $F_{x}^{(3)}$  & $\delta_x$ & $ f_{o/x}^{(4)}$ \\
\hline
\endfirsthead
\caption{(continued)}\\
\hline\hline
Events  & $N_H^{(1)}$ & $F_{o}^{(2)}$   & $\delta_o$ & $F_{x}^{(3)}$  & $\delta_x$ & $ f_{o/x}^{(4)}$ \\
\hline
\endhead
\hline
\endfoot
\object{XRF 981226}  & 1.78&   $<1.7$  & & $4.88^{+0.24}_{-0.44}$$^{(5)}$ & $-(0.66^{+0.68}_{-0.44})$$^{(8)}$  & $<0.34$ \\
\object{XRF 990520}   & 3.6&   $ <5.7$  &  & & & \\
\object{XRF 990704}   & 3.0&   $<1.2$  & & $5.95\pm0.78$$^{(5)}$ & $ -(0.88^{+0.28}_{-0.20})$$^{(8)}$ & $<0.20$ \\
\object{XRR 991106}  & 46&   $ <34$  & & $2.10\pm0.65$$^{(5)}$ & $ -(1.1^{+2.5}_{-2.1})$$^{(8)}$  & $ <16$ \\
\object{XRF 000416}  & 8.0&   $ <150$  & &  &  &  \\
\object{XRF 000615}  & 2.7&   $<2.7$  & & $1.28\pm0.20$$^{(5)}$ & $ (0.23^{+1.32}_{-0.94})$ $^{(8)}$ & $<2.1$ \\
\object{XRF 010213}  & 3.4&   $ <27$  & &  &  &  \\
\object{XRR 010326B} & 4.3&   $ <15$  & &  &  &  \\
\object{XRR 010629} & 14&   $ <36$  & &  &  &  \\
\object{XRR 010921} & 19&   $ <71$  & &  &  &  \\
         &    &  $210\pm40$ & -1.6$^{(10)}$ & & & \\
\object{XRF 011019}  & 2.6&   $ <8.3$  & &  &  &  \\
\object{XRR 011030}  & 10&   $ <11$  & & $30.8\pm0.9$ $^{*}$ &   & $ <0.39$ \\
\object{XRF 011130}  & 3.3&   $ 70\pm11$  & -1.30$^{+0.04}_{-0.05}$$^{(11)}$ & $0.99\pm0.92$$^{*}$ &   & $ 71\pm66$ \\
\object{XRR 011211}   & 4.3&   $ 23.6\pm0.87$  & $ -0.96\pm0.06$$^{(12)}$ & $1.0\pm0.1$$^{(5)}$ & $-(1.6\pm0.3)$ $^{(13)}$ & $24\pm3$ \\
\object{XRF 011212}    & 34 &   $ <8.3$  & &  &  &  \\
\object{XRR 020124}   & 4.1   &  $4.03\pm0.52$ & & & & \\
\object{XRR 020127}    & 5&   $ <20$  & & $7.1\pm1.9$$^{*}$ &  & $<2.8$ \\
\object{XRF 020317}    & 3.8   &  $85.4\pm7.8$ & & & & \\
\object{XRR 020410}    & 6.4  &  $ 8.8\pm4.1$  &$<-0.95$$^{(14)}$ & $77.8^{+6.3}_{-6.9}$$^{(5)}$ & $-(0.92\pm0.12)$$^{(8)} $  & $ 0.11\pm0.05$ \\
\object{XRF 020427}    & &    & &  $5\pm2.5$$^{(5)}$ & $-(1.3^{+0.10}_{-0.12})$$^{(8)} $  &     \\
\object{XRF 020625}    & 7.9  &  $<200$ & & & &  \\
\object{XRR 020812}      & 4  &  $<20$ & & & & \\
\object{XRR 020819}       & 5.4  &  $<13$ & & & & \\
\object{XRF 020903}     & 2.3&   $ (12\pm2)\times10^{3}$  & & & &  \\
                    & &  $<1.9$ & & & & \\
\object{XRR 021004} & 4.3  &  $140\pm7$ & -1.3 $^{(15)}$& $ 14\pm1$$^{**}$ &$-(1.0\pm0.2)$$^{(16)}$  & 10$\pm1$ \\
\object{XRR 021104}           & 21  &  $<2.5$ & & & & \\
\object{XRR 021112}          & 15  &  $ <3.1$ & & & & \\
\object{XRR 021211}          & 3.7  &  $3.11\pm0.53$ & $-(0.96\pm0.04)$$^{(17)}$ & & & \\
\object{XRR 030115}             & 1.9  &  $<13$ & & & & \\
\object{XRR 030323}             & 4.8  &  $75.9\pm0.63$ & & & & \\
\object{XRR 030324}            & 1.9 &  $<7$ & & & & \\
\object{XRR 030329}    & 2.2&   $ 3800\pm100$  &$-0.97\pm0.03$$^{(18)}$ & $500\pm50$$^{**}$ & -($0.9)$$^{(19)}$   & $ 7.64\pm0.81$ \\
\object{XRF  030416}           & 4.6  &  $<4.1\times10^{3}$ & & & & \\
\object{XRR 030418}         & 3.3  &  $4.8\pm3.4$ & & & & \\
\object{XRF 030429}         & 5.1  &  $23.6\pm3.3$ & & & & \\
\object{XRF 030528}   & 15&   $ <0.4$  & & $5.3\pm2.4$$^{*}$ &  & $ <0.07$ \\
\object{XRF 030723}    & 1.5&   $ 4.9\pm1.8$  &$-0.10\pm0.06$$^{(20)}$ & $ 5.9\pm1.3$$^{*}$  &  & $ 0.83\pm0.35$ \\
\object{XRR 030725}          & 3.3  &  $75\pm15$ &-0.9$^{(21)}$ & & & \\
\object{XRF 030823}             & 8.2  &  $<7.5$ & & & & \\
\object{XRF 030824}            & 4.1  &  $<14$ & & & & \\
\object{XRR 031220}   & 11&   $ <8.7$  & & $2.2\pm0.8$$^{*}$ &  & $<4.5$ \\

\end{longtable}
(1){\footnotesize:in units of $1\times10^{20}cm^{-2}$,(2)= in units of  $\mu$J, (3)=in units of  $10^{-13}erg\,cm^{-2}\,s^{-1}$,  (4)=optical flux in units of  $\mu$J and X-ray  flux in units  of $10^{-13}erg\,cm^{-2}\,s^{-1}$, (5)=\citet{depasq5}, (6)=\citet{fro00}, (7)=\citet{fero01},(8)=\citet{depasq5} (9)=\citet{Miao04}, (10)=\citet{priceb}, (11)=\citet{gar}, (12)=\citet{ses}, (13)=\citet{piro04}, (14)=\citet{nic04}, (15)=\citet{weid02},  (16)=\citet{rsak1}, (17)=\citet{cor1}, (18)=\citet{price3}, (19)=\citet{tiego},  (20)=\citet{fynbo}, (21)=\citet{vin},*=flux extrapolated using the decay index between the  prompt data and the first ToO observations, **= flux extrapolated using the early afterglow decay index from the  observation nearest to $40$ $Ks$.}

\newpage



\newpage

\begin{table*}[t]
\caption{ Intrinsic mean  value  and  variance of  the  logarithm of the X-ray [1.6-10 \kev] and optical flux [R band], at 40 $ks$  for
XRRs/XRFs  with early afterglow observations  and    GRBs. The X-ray flux is in units  of $erg\,cm^{-2}\,s^{-1}$ and the optical flux
in units  of $W\,m^{-2}\,Hz^{-1}$.  }
\label{tab12}
\begin{tabular}{|l|cc|cc|}
\hline
Classes & $ \langle log(f_{o11.1})\rangle_i $ & $ \langle \sigma \rangle_i$  & $\langle log(f_{x11.1}) \rangle_i$ & $ \langle \sigma
\rangle_i$  \\
\hline
XRRs/XRFs               &  &   &  -$(12.07^{+0.48}_{-0.52})$ &0.85$^{+0.46}_{-0.30}$    \\
XRRs/XRFs OT             &   -($30.38^{+0.48}_{-0.44})$ &   0.86$^{+0.26}_{-0.14}$ &   &                      \\
XRRs/XRFs without  known redshift        &   &        & -$(12.24^{+0.30}_{-0.27})$&  $0.51^{+0.22}_{-0.11}$      \\
GRB                        &  &   &-$(12.12^{+0.16}_{-0.14})$   &0.45$^{+0.14}_{-0.12}$\\
OTGRB                  & -$(30.54\pm0.32)$ &    0.53$^{+0.21}_{-0.13}$  & &     \\

\hline
\multicolumn{5}{c} { }
\end{tabular}
\end{table*}

\end{document}